\documentclass[10pt,twocolumn,letterpaper]{article}

\usepackage{cvpr}              % To produce the CAMERA-READY version
\usepackage{float} % Add this line in the preamble
\definecolor{cvprblue}{rgb}{0.21,0.49,0.74}
\usepackage[pagebackref,breaklinks,colorlinks,allcolors=cvprblue]{hyperref}

\usepackage{afterpage}

\usepackage[linesnumbered, ruled, vlined]{algorithm2e}
\SetKwInput{KwInput}{Input}                % Set the Input
\SetKwInput{KwOutput}{Output}              % set the Output

\def\paperID{17076} % *** Enter the Paper ID here
\def\confName{CVPR}
\def\confYear{2025}

\newcommand{\name}{PEFTLeak}
\newcommand{\namespace}{PEFTLeak }

\title{Gradient Inversion Attacks on  Parameter-Efficient Fine-Tuning}

\author{
    Hasin Us Sami, Swapneel Sen, Amit K. Roy-Chowdhury, Srikanth V. Krishnamurthy,  Ba\c{s}ak G\"{u}ler \\
    University of California, Riverside, CA \\
    {\tt\small hsami003@ucr.edu, ssen010@ucr.edu, amitrc@ece.ucr.edu, krish@cs.ucr.edu, bguler@ece.ucr.edu}
}

\begin{document}
\maketitle
\begin{abstract}
Federated learning (FL) allows multiple data-owners to collaboratively train machine learning models by exchanging local gradients, while keeping their private data on-device. To simultaneously enhance privacy and training efficiency, recently parameter-efficient fine-tuning (PEFT) of large-scale pretrained models has gained substantial attention in FL. While keeping a  pretrained (backbone) model frozen, each user fine-tunes only a few lightweight modules to be used in conjunction,  to fit specific downstream applications. Accordingly, only the gradients with respect to these lightweight modules are shared with the server. In this work, we investigate how the privacy of the fine-tuning data of the users can be compromised via a  malicious design of the pretrained model and trainable adapter modules. We demonstrate gradient inversion attacks on a popular PEFT mechanism, the adapter, which allow an attacker to reconstruct local data samples of a target user, using only the accessible adapter gradients. 
Via extensive experiments, we demonstrate that a large batch of fine-tuning images can be retrieved with high fidelity. 
Our attack highlights the need for privacy-preserving mechanisms for PEFT, while opening up several future directions. 
Our code is available at \url{https://github.com/info-ucr/PEFTLeak}. \end{abstract}
%\url{https://github.com/PEFTLeak/Inversion}.

\section{Introduction}
Federated learning (FL) is a collaborative training paradigm to train a machine learning model across multiple data-owners (users)  \cite{mcmahan2016communication}. Users perform training locally using their local data and send the local model updates/gradients to a central server. The server aggregates these to form a global model. 
By obviating the need to share raw local samples, FL has emerged as a promising framework in fields where data privacy is paramount such as healthcare.

Recently, leveraging large-scale pretrained models in various downstream tasks has gained significant attention, owing to their remarkable training performance. Considering the potential of pretrained models, recent works have extended their application to FL \cite{weller2022, Qu2022, John2023}. However, these pretrained models often contain a massive number of parameters, which can be in the range of millions/billions. Full fine-tuning (FFT) of such large models and communicating the gradient parameters require often prohibitive computational infrastructure and bandwidth. This prevents users with low computation/communication resources from participating in training and  potentially causing bias in the global model. Thus, recent works have explored parameter-efficient fine-tuning (PEFT) \cite{Houlsby2019, Pfeiffer2020, Lisa2021, Edward2022, Shoufa2022, Dong2023, Wei2024, Imad2024}, where in lieu of fine-tuning the entire pretrained model, only a small number of lightweight modules are trained; the backbone model is kept frozen. Due to marked reduction in resource consumption and training latency, PEFT has become widely popular in  FL   \cite{Yeachan2023, Zhang2023, Mingchen2023, Sun2024, Malaviya2023, Chulin2024}.

 Though FL is popular in privacy-sensitive tasks, an adversarial server can still reverse-engineer the local gradients received from the users to extract privacy-sensitive information about local data samples \cite{Zhu2019, Nasr2019, Geiping2020, Yin2021, Yangsibo2021, Jeon2021, Ali2022, Lu2022, Wen2022, Francisco2023, Vu2024, Zhang2024}. These attacks can be grouped into: 1) gradient inversion attacks \cite{Geiping2020, Yin2021, Ali2022, Fowl2022} and, 2) membership inference attacks \cite{Shokri2017, Nasr2019, Vu2024, Wen2024}.  Gradient inversion attacks reconstruct the raw data samples held by a target user using the gradient received from the user.  In contrast, membership inference attacks seek  whether a  candidate sample  belongs to the local dataset of a  target user. 
 The success of membership inference attacks under both FFT and PEFT is shown  via a malicious design of the pretrained model \cite{Wen2024, Liu2024}.  These works do not consider gradient inversion attacks, which can be more detrimental to privacy, as they can operate without the knowledge of candidate samples  \cite{Feng2024}.   
 
 % In this work, we study gradient inversion attacks to PEFT. 
 A gradient inversion attack is proposed in \cite{Feng2024}, by poisoning the pretrained vision transformer (ViT) parameters \cite{Dosovitskiy2021}. 
However, this attack relies on FFT and assumes that the attacker has access to the gradients corresponding to the entire model. Such attacks do not apply to PEFT methods, where the pretrained model remains frozen and the attacker loses access to the full (local) gradient. Due to this, PEFT has been considered to  enhance privacy and limit the risk of exposing sensitive information   \cite{Zhang2023};  it has been shown  that conventional gradient inversion attacks \cite{Zhu2019} have a reduced success rate under PEFT. However, these attacks do not consider more powerful adversaries where the attacker can deviate from the benign training protocol to breach privacy,  such as changing the model architecture and/or parameters sent to the users  \cite{Fowl2022, Fowl2023, Feng2024}.

\begin{figure}
  \centering
   \includegraphics[width=0.7\linewidth]{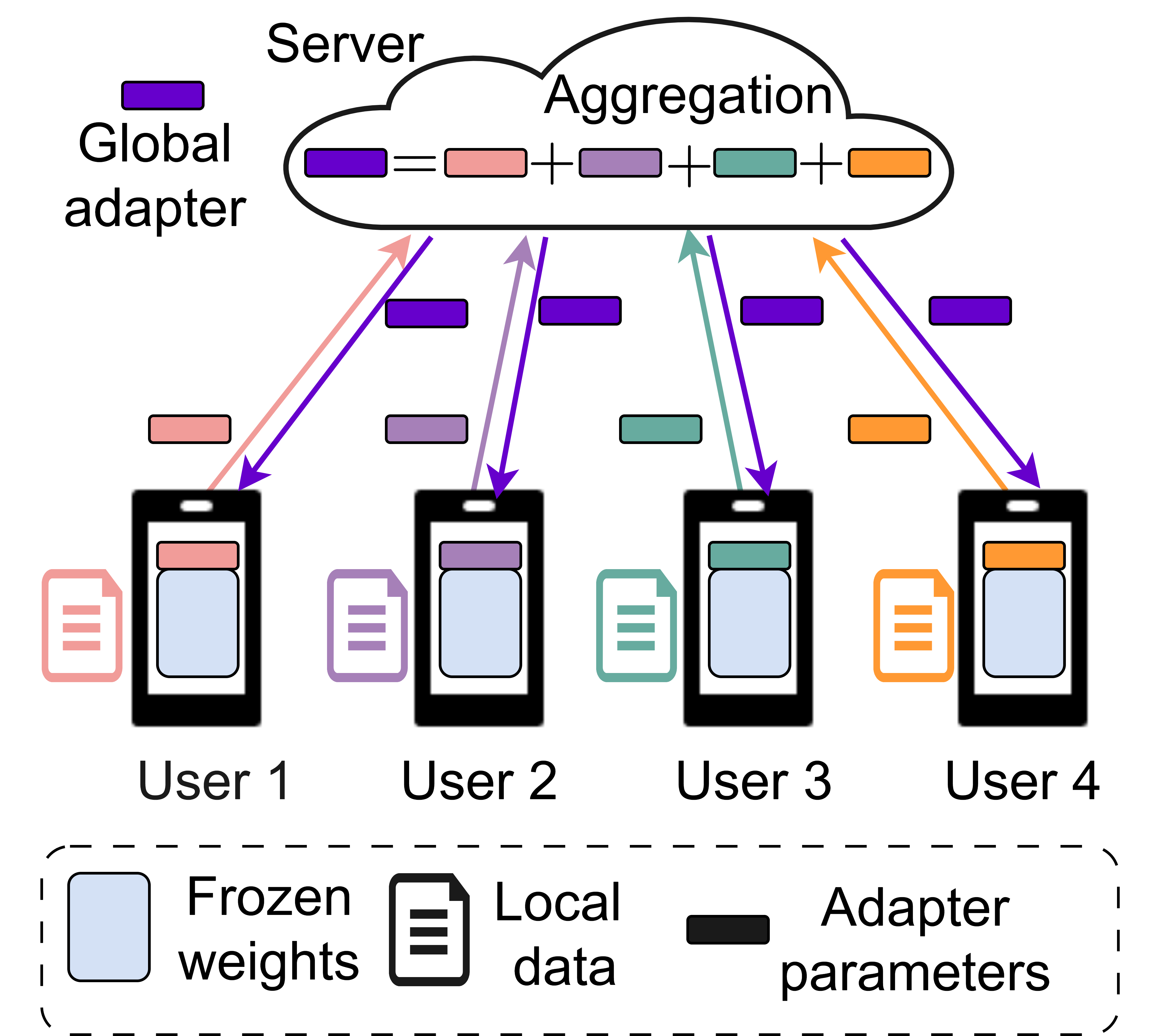}
   %\vspace{-1cm}
 \vspace{-0.05cm}  \caption{A small scale setup of PEFT based FL with $U=4$ users. In each training round, users fine-tune the lightweight adapter modules and send the adapter parameters to the server. The backbone model parameters are kept frozen. After aggregation, the server sends the global  adapter parameters back to the users. } 
   \label{fig:FL_adapter}
\vspace{-0.35cm}\end{figure}

In this work, we ask the question whether gradient inversion attacks are viable for PEFT. Specifically, we consider  
% Towards analyzing the robustness of PEFT against more capable adversaries, 
gradient inversion attacks on a popular PEFT mechanism, \emph{adapters}, proposed in  \cite{Houlsby2019}.  
This method introduces lightweight modules, called \emph{adapters}, inside the backbone model. Originally introduced for natural language processing tasks, adapters gained significant popularity in vision tasks recently \cite{Sung2022, Mingchen2023, Imad2024}.
In an FL setup, each user computes gradients with respect to these adapter parameters only and sends them to the server (attacker), while the backbone model is frozen and not updated during training \cite{Mingchen2023, Zhang2023, Yeachan2023}. The server computes an average of local adapter gradients to form the global (aggregated) adapter parameters. These global parameters are sent back to the users for the next training round. 
The adversary seeks access to the local image dataset used for local fine-tuning by the users. 
Compared to conventional gradient inversion attacks targeted at FFT, a major limitation in the PEFT setting (from the attacker's perspective) is that the attacker can only observe the local gradients for a small number of adapter parameters, as the frozen backbone model does not carry any new information.  
In this context, whether an attacker can recover sensitive training samples of a target user (victim) using the limited information provided by the adapter gradients is yet unknown.

In response to the question above, we propose a novel inversion attack, \name, that can successfully recover the data samples of the victim  {\em by only leveraging} the gradients from the adapters. 
Our attack builds on poisoning the pretrained model and adapter modules, in a way to  
lead the victim to unintentionally leave a footprint of their raw data inside the adapter gradients. To the best of our knowledge, our work is the first to propose a successful inversion attack applicable to a PEFT method. 
% , on the adapter setting. 
Our attack highlights the need for stronger defenses for adapters, and provides a number of principles that can also be useful for investigating inversion attacks for other PEFT methods like prefix-tuning \cite{Lisa2021}, bias-tuning \cite{Han2020}, or low-rank adaptation  \cite{Edward2022}. 

Our contributions are summarized as follows:
\begin{enumerate}
\item We  demonstrate the first successful gradient inversion attack on PEFT.  Our attack  can successfully extract local data samples of a victim user using the gradients of lightweight adapters, as opposed to the full model.

\item Extensive experiments on CIFAR-10, CIFAR-100 \cite{krizhevsky2009learning} and TinyImageNet \cite{Tinyimagenet} demonstrate that the attacker can breach privacy of fine-tuning data with high-accuracy. 

% \krish{Perhaps a third contribution can be a discussion of defenses?}

% \basak{Added a discussion of defenses below. }

\item We further show that reducing the adapter dimension (number of parameters) does not guarantee privacy. For small dimensions our attack can be executed in multiple rounds to increase the number of reconstructed patches.

\item  Our results suggest that PEFT mechanisms, though greatly reducing the number of parameters shared during training, should not be viewed as a defense against inversion attacks, and highlights the necessity of stronger defenses such as differentially private PEFT for FL.  
\end{enumerate} 
\section{Related Works}
\noindent
{\bf PEFT.}  
PEFT mechanisms combat large computation and memory costs associated with fine-tuning the entire pretrained model. Such methods freeze the pretrained model while tuning only the bias parameters \cite{Han2020}, introducing additional lightweight trainable modules \cite{Houlsby2019, Pfeiffer2020, Shoufa2022, Dong2023, Imad2024}, learnable tokens \cite{Lisa2021, Brian2021, Menglin2022} or optimizing low-rank matrices \cite{Edward2022, Yang2023, Yeming2024}. To enhance training efficiency, recent works have extended PEFT to FL \cite{Aliaksandra2023, Zhang2023, Mingchen2023, Chulin2024}.  

\noindent
{\bf Privacy attacks.} Gradient inversion attacks retrieve local data samples using local gradients. Attacks from \cite{Zhu2019, Geiping2020} recover ground-truth images by minimizing the distance between the true and estimated gradients. Subsequent works leverage batch-normalization statistics \cite{Yin2021}, blind source separation \cite{Sanjay2023} or a pretrained generative model \cite{Jeon2021, Li2022, Hao2023} to recover a batch of images. These works consider fully connected or convolutional neural networks. Attacks on  ViTs are explored under the FFT setting in \cite{Ali2022, Lu2022, Daniel2023, Zhang2024}.   
References \cite{Fowl2022, Fowl2023, Boenisch2023, Joshua2023, Hong2023, LOKI2024, Feng2024} consider more powerful adversaries who can manipulate the model parameters/architecture to further improve reconstruction. 

%\noindent
%{\bf Membership Inference Attack. } 
Membership inference attacks seek whether a specific data sample belongs to the victim user \cite{Shokri2017, Nasr2019, Song2021}. Such attacks have been explored under both FFT and PEFT settings \cite{Fu2023, Vu2024, Wen2024, Abascal2024}.  
Other works recover personally identifiable information or infer the presence of a target property \cite{Carlini2021,Lukas2023, Tian2023, Xiaoyi2023, Liu2024}. 
%These attacks are complementary to our attack.  
In contrast, our goal in is to recover the fine-tuning dataset held by the victim user.

\section{Problem Formulation}

We consider a typical centralized FL framework  with $U$ users. Training is controlled by a server -often an organization with abundant resources-  who performs pretraining using proprietary data or publicly available proxy datasets. Fine-tuning enhances task-specific performance by leveraging data from end-users. User $i$ holds a local dataset $\mathcal{D}_i$. At each training round, user $i$ downloads the current state of the global adapter parameters $\mathbf{w}_{A} \in \mathbb{R}^d$ from the server and performs training locally using  its on-device dataset. The goal is to train $\mathbf{w}_A$ to minimize  the global loss,
\vspace{-0.2cm}\begin{equation}
   \mathcal{L}(\mathbf{w}_F, \mathbf{w}_A) \triangleq \frac{1}{U} \sum_{i \in [U]}\mathcal{L}_i(\mathbf{w}_F,\mathbf{w}_A)
\vspace{-0.15cm}\end{equation}
where $\mathcal{L}_i$ is the local loss of user $i$ and $\mathbf{w}_F$ is the frozen pretrained model. The frozen parameters are used in forward propagation to compute the loss, but are not updated during backpropagation. 
This system is illustrated in Fig. \ref{fig:FL_adapter}. 

\noindent
{\bf Threat Model.} We consider a malicious server who can modify the pretrained model  and global  adapter parameters. 
The server sends the pretrained model to the users once prior to training, after which this model is frozen (not updated), hence does not carry any new information from the users.   
The server sends the global adapter parameters to the users at each training round, which are then updated by the users.  
After receiving the local adapter gradients, the server initiates the attack to recover the data belonging to the victim user. 
% The attack can be executed within a single training round. 
Our threat model is motivated by \cite{Fowl2022, Fowl2023, Hong2023, Feng2024}.

\begin{figure}
  \centering
   \includegraphics[width=0.75\linewidth]{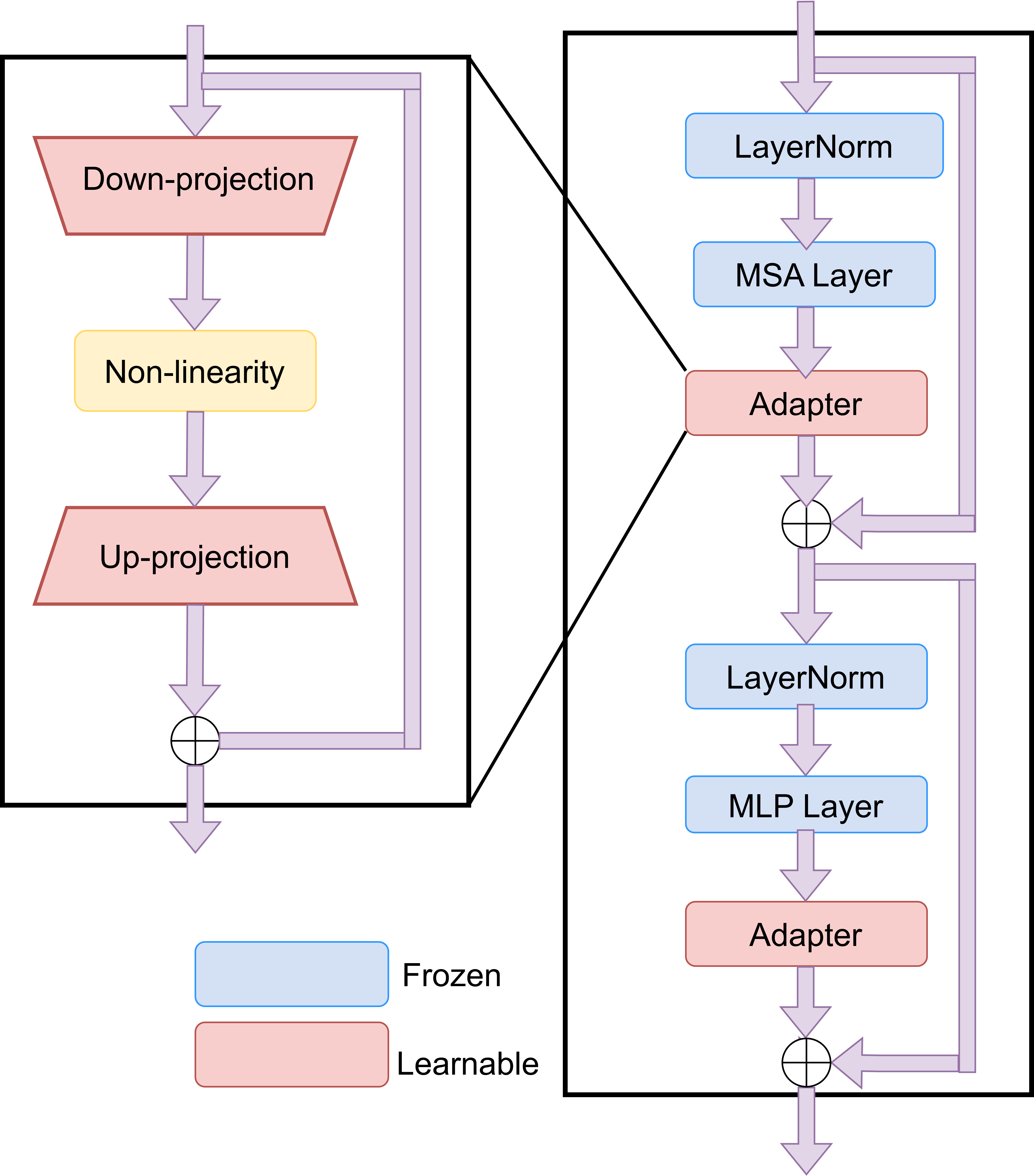}
   %\vspace{-1cm}
   \vspace{-0.2cm} \caption{{\bf ViT encoder with adapter modules.} ViT encoder with stacked LayerNorm, MSA, MLP layers and residual connections. An adapter module is inserted after each MSA/MLP layer, consisting of two feed-forward blocks and a non-linearity in-between.  } 
   \label{fig:ViT}
\vspace{-0.4cm}\end{figure}

\noindent
{\bf ViT with adapters. } In ViT, an image is split into multiple patches. Sequences of patch embeddings are generated first and then a distinct position encoding vector is added to each patch embedding. Next, these position-encoded embeddings are used as inputs to multiple stacked multi-head self-attention (MSA) and multi-layer perceptron (MLP) layers. Each MLP layer consists of two fully connected (FCN) layers with a GELU \cite{Hendrycks2016} activation function in-between. In addition, a class token is used to predict classes of the fine-tuning samples at the final classification layer (also known as classification head). For fine-tuning, an adapter block is inserted after each MSA and MLP layer \cite{Houlsby2019, Mingchen2023, Imad2024} as shown in Fig.~\ref{fig:ViT}. Each adapter block consists of two FCN layers. The first layer projects the original embeddings of dimension $D$ into a lower dimension $r$ ($r<<D$), followed by a non-linear activation function. The second layer maps the projected dimension back to the original dimension, $D$. By choosing $r<<D$, the number of trainable parameters is significantly reduced in PEFT compared to FFT.

The most relevant work to ours is the recent inversion attack to FFT \cite{Feng2024}, where users send full transformer gradients to the server (as opposed to PEFT).  
The attacker maliciously modifies the transformer parameters to recover sensitive local training   images.  
Let us denote the total number of patches as $N$, images in the batch as $M$, and the  (sensitive) patch embeddings by $\{\mathbf{y}^{(n,m)}\}_{n \in [N], m \in [M]}$, which is the sum of image patch embedding and position encoding vectors. 
The class token embedding is denoted by $\mathbf{y}^{(0,m)}$. 
The attack 
%uses a malicious design of transformer parameters to 
allows uninterrupted flow of $\mathbf{y}^{(n,m)}$ until the MLP layer.   
Let  $\mathbf{w}_j$ and $b_j$, 
denote the weight vector and bias corresponding to neuron $j$ in the first FCN layer. 
The output for neuron $j$ can be written as, 
\begin{equation} \label{eq:neuron_output}
    v_j^{(n,m)}\hspace{-0.05cm} \triangleq \mathbf{w}_j^{\text{T}}\mathbf{y}^{(n,m)}\hspace{-0.05cm}+\hspace{-0.03cm}b_j \hspace{0.1cm} \forall n \in \{0, \ldots, N\}, m \in [M]
\end{equation}  
where the  gradient of user $i$ with respect to $\mathbf{w}_j$ and $b_j$ is,
\begin{align} 
   \frac{\partial \mathcal{L}_i} {\partial \mathbf{w}_j} 
   % & \triangleq \frac{1}{M(N+1)}\sum_{m=1}^{M}\sum_{n=0}^{N} \frac{\partial \mathcal{L}_i} {\partial v_j^{(n,m)}}\frac{\partial v_j^{(n,m)}} {\partial \mathbf{w}_j} \notag\\
   & = \frac{1}{M(N+1)}\sum_{m=1}^{M}\sum_{n=0}^{N} \frac{\partial \mathcal{L}_i} {\partial v_j^{(n,m)}}\mathbf{y}^{(n,m)} \label{eq:weight_grad} \\ 
   \frac{\partial \mathcal{L}_i} {\partial b_j} 
   % & \triangleq \frac{1}{M(N+1)}\sum_{m=1}^{M}\sum_{n=0}^{N} \frac{\partial \mathcal{L}_i} {\partial v_j^{(n,m)}}\frac{\partial v_j^{(n,m)}} {\partial b_j}\notag\\
   & =  \frac{1}{M(N+1)}\sum_{m=1}^{M} \sum_{n=0}^{N} \frac{\partial \mathcal{L}_i} {\partial v_j^{(n,m)}} \label{eq:bias_grad}
\end{align}
which represents the average of gradients for all patches and class tokens across the images. 
Let us assume only patch $n$ from image $m$ propagates through the activation function in neuron $j$ while all other patches are blocked. 
Then,
\begin{align} 
   \frac{\partial \mathcal{L}_i} {\partial \mathbf{w}_j} &=  \frac{1}{M(N+1)}\frac{\partial \mathcal{L}_i} {\partial v_j^{(n,m)}}\mathbf{y}^{(n,m)}, \notag\\
    \quad 
   \frac{\partial \mathcal{L}_i} {\partial b_j} &= \frac{1}{M(N+1)}\ \frac{\partial \mathcal{L}_i} {\partial v_j^{(n,m)}} \notag 
\end{align} 
and embedding $\mathbf{y}^{(n,m)}$ can be recovered as,
\begin{equation} \label{eq:base_attack}
    {\frac{\partial \mathcal{L}_i} {\partial \mathbf{w}_j}} \Big / { \frac{\partial \mathcal{L}_i} {\partial b_j} }=\mathbf{y}^{(n,m)} 
\end{equation}   
The attacker then designs the weight and bias parameters in the MLP layer with the constraint that no two patches/tokens activate the same set of neurons, by leveraging patch statistics obtained from a public dataset which was originally proposed in \cite{Fowl2022}.  
After recovering the embeddings  $\mathbf{y}^{(n,m)}$, one can recover the image patch embeddings by subtracting the position encoding vectors from $\mathbf{y}^{(n,m)}$. 

\noindent
{\bf Challenges. } In adapter-based PEFT  \cite{Houlsby2019}, MLP layers are  initialized prior to training and kept frozen afterwards (not updated by users). As a result, these layers do not carry any information about the local fine-tuning data, making the aforementioned attack infeasible.   
A  potential approach is then to use the adapter gradients to recover the image embeddings as in \eqref{eq:base_attack}. 
For this, the target embeddings $\{\mathbf{y}^{(n,m)}\}_{n \in [N], m \in [M]}$ should propagate through all intermediate layers before the adapter without any significant shift in the original contents, which requires the frozen MSA and MLP layers act as identity mappings. 
The key challenge is then the projection to lower dimension in the adapter, due to which the attacker has access to a maximum of $r$ weight and bias gradients. As seen in \cite{Houlsby2019,Shoufa2022}, the typical values for $r$ lie within the range of $1$ to $64$. As evident in \eqref{eq:base_attack}, a single neuron's weight and bias gradients play a role in recovering a single image patch. Reconstruction rate increases with the number of available neurons, which was also shown in \cite{Fowl2022, LOKI2024}.  
The MLP layer of the ViT architecture from \cite{Dosovitskiy2021} has $3072$ neurons.  These $3072$ neurons' gradients were leveraged for the attack in \cite{Feng2024}, leading to the succesful recovery of a large batch of images. 
% In contrast, the recovery rate from only the $r$ neurons within a single adapter will challenge the recovery of a large batch of images. 
In contrast, fine-tuning results in modifying only a small set of parameters (corresponding to $r\leq 64$ neurons in the adapter)  compared to the  images, which makes it hard to reconstruct a large fraction  of the images using this limited information.

\noindent
{\bf Contributions. } We address these challenges by crafting a novel gradient inversion attack on  adapter-based PEFT. 
To do so, we introduce a malicious design of the pretrained model and adapters, which can uncover a large number of private fine-tuning images with a much reduced observable space  compared to what was available with FFT. 
Our attack  tackles the dimensionality problem (limited number of neurons) for the adapter by leveraging the gradients from multiple adapter layers, and carefully designs the patch propagation structure to enable different adapter layers to recover images with different features.

\section{Framework} \label{sec:framework}

We next describe the details of \name. 
We consider the conventional ViT architecture from \cite{Dosovitskiy2021} as the pretrained model. The architecture along with inserted adapter modules is depicted in Fig. \ref{fig:ViT}. LayerNorm, MSA and MLP layers are kept frozen while only adapter modules are trainable \cite{Shoufa2022, Imad2024}. We assume that the victim performs fine-tuning on a batch of $M$ images. We denote the $m^{th}$ image in the batch as $\mathbf{X}(m) \in \mathbb{R}^{C\times H \times W}$ for $m \in [M]$, where $C$ is the number of channels, and $(H,W)$ denotes the image resolution.  
Each image is divided into $N$ patches of resolution $(P,P)$. Next, each patch is flattened to a vector, and denoted as $\mathbf{x}^{(n, m)} \in \mathbb{R}^{P^2C}$ for $n \in [N]$, $m \in [M]$.  
We assume  that the range of input values in $\mathbf{x}^{(n,m)}$ are in $[-1,1]$  \cite{Pasquini2022, LOKI2024}.
The patches are then mapped to a dimension $D$ through linear projection using the weight matrix $\mathbf{E} \in \mathbb{R}^{D \times P^2C}$ from the pretrained model, 
\begin{equation} \label{eq:x_map}
    \mathbf{x}_{map}^{(n,m)} \triangleq \mathbf{E} \mathbf{x}^{(n,m)}
\end{equation}
for $n \in [N]$. 
A class token $\mathbf{x}_{map}^{(0,m)}$ is used  to predict the class of  image $m$ in the classification layer, which  
 propagates like the image patch embeddings throughout the intermediate layers.   
Next, position encoding vectors $\mathbf{E}_{pos}^{(n)} \in \mathbb{R}^D$ for $n \in \{0, \ldots, N\}$ are added to the embeddings, 
% to encode position information,
\begin{equation} \label{eq:y}
    \mathbf{y}^{(n,m)} \triangleq  \mathbf{x}_{map}^{(n,m)}+\mathbf{E}_{pos}^{(n)}= \mathbf{E} \mathbf{x}^{(n,m)}+ \mathbf{E}_{pos}^{(n)}
\end{equation}

\noindent
{\bf Key intuition. } At a high-level, our goal is to propagate the embeddings $\mathbf{y}^{(n,m)}$ up to the adapter layers without any significant distortion.  
This should be achieved  by a one-time adversarial modification of the pretrained model, which is sent to the users prior to fine-tuning. As these layers are frozen and not modified by the users during fine-tuning, they provide no information about the local image patches. 
Adapter layers will then be utilized to recover the image patches. These trainable parameters are modified by the users during fine-tuning, and hence encode sensitive information about the local images. On the other hand, the dimensionality of a single adapter layer is not sufficient to recover a large batch of images. To that end, our attack is designed across multiple adapter layers, where different layers capture image patches with varying statistics.  
We next describe the design of the individual layers.

\subsection{First LayerNorm (LN1) } \label{sec:LN1}

%\vspace{0.1cm}
%\noindent
%{\bf 1. First LayerNorm (LN1). }

 After position encoding, the embeddings  $\mathbf{y}^{(n,m)}$ enter a LayerNorm (LN1) layer, 
\begin{equation} \label{eq:LN1}
    \mathbf{z}^{(n,m)} \triangleq \frac{\mathbf{y}^{(n,m)}- \mu^{(n,m)}}{\sigma^{(n,m)}} \odot \mathbf{w}_{LN1} + \mathbf{b}_{LN1}
\end{equation}
where $\mathbf{w}_{LN1}, \mathbf{b}_{LN1} \in \mathbf{R}^D $ are the weight and bias parameters at LN1. 
$\mu^{(n,m)}, \sigma^{(n,m)}$ denote the mean and standard deviation across the elements in $\mathbf{y}^{(n,m)}$ and $\odot$ denotes element-wise multiplication. 
As our goal is to ensure uninterrupted flow of information from the input layer, we want $\mathbf{z}^{(n,m)} \approx \mathbf{y}^{(n,m)}$. 
For this, we design $\mathbf{E}$ to make the mean and standard deviation across $\mathbf{x}_{map}^{(n,m)}$ in \eqref{eq:x_map} negligible for all $n \in [N], m \in [M]$ compared to the mean and standard deviation of $\mathbf{E}_{pos}^{(n)}$. 
Selecting $ \mathbf{E}_{pos}^{(n)}\sim \mathcal{N}(0,\sigma)$ 
 with $\sigma=10$ and $\mathbf{E}=0.5\mathbf{I}_D$ satisfies these conditions, 
$\mu^{(n,m)} \approx 0$, and $\sigma^{(n,m)} \approx \sigma$ for all $n \in \{0, \ldots, N\}$.  
Then, by setting each element in $\mathbf{w}_{LN1}$ to $\sigma$ and $\mathbf{b}_{LN1}$ to $0$, $\mathbf{z}^{(n,m)} \approx \mathbf{y}^{(n,m)}$.

%\noindent
%{\bf Multi-head Self Attention (MSA). } 

\subsection{Multi-head Self Attention (MSA)} \label{sec:MSA}

The output embeddings of LN1 enter the Multi-head Self Attention (MSA) layer. 
To enable undistorted propagation of input patch embeddings, we design the weight parameters in the MSA layer to act as identity mappings. 
% as follows.  
Suppose that the MSA layer consists of $L$ heads where the embedding dimension of each head is  $D_h \triangleq D/L$. For head $h$,  denote the query, key and value weight matrices as $\mathbf{W}_Q^h, \mathbf{W}_K^h$ and $\mathbf{W}_V^h$, and biases  as $\mathbf{b}_Q^h, \mathbf{b}_K^h$ and $\mathbf{b}_V^h$, respectively. 
Let,   
\vspace{-0.05cm}\begin{align} \mathbf{W}_Q^h&=\mathbf{W}_K^h=\mathbf{W}_V^h= \mathbf{I}_{D_h \times D_h}  \label{eq:weight_MSA}\\
    &\mathbf{b}_Q^h= \mathbf{b}_K^h= \mathbf{b}_V^h=\mathbf{0} \label{eq:bias_MSA}  
\end{align} 
and define $(\mathbf{y}^{(n,m)})_h \triangleq \mathbf{y}^{(n,m)}[hD_h:(h+1)D_h] \in \mathbb{R}^{D_h}$ 
as the $D_h$ elements from the $n^{th}$ patch embedding  that propagates through head $h$. 
Then, query, key and value matrices for head $h$ are given by,
\vspace{-0.05cm}\begin{align}
    \mathbf{Q}^h &\triangleq \begin{bmatrix}\mathbf{W}_Q^h (\mathbf{y}^{(0,m)})_h &\cdots &\mathbf{W}_Q^h (\mathbf{y}^{(N,m)})_h \end{bmatrix}\notag\\
    % &=\begin{bmatrix}(\mathbf{y}^{(0,m)})_h &\cdots &\ (\mathbf{y}^{(N,m)})_h \end{bmatrix} \notag \\
    \mathbf{K}^h &\triangleq \begin{bmatrix}\mathbf{W}_K^h (\mathbf{y}^{(0,m)})_h &\cdots &\mathbf{W}_K^h (\mathbf{y}^{(N,m)})_h \end{bmatrix}\notag\\ 
    % &=\begin{bmatrix}(\mathbf{y}^{(0,m)})_h &\cdots &\ (\mathbf{y}^{(N,m)})_h \end{bmatrix} \notag \\ 
    \mathbf{V}^h &\triangleq \begin{bmatrix}\mathbf{W}_V^h (\mathbf{y}^{(0,m)})_h &\cdots &\mathbf{W}_V^h (\mathbf{y}^{(N,m)})_h \end{bmatrix}  
\end{align} 
%Note that for $n \neq t$, the position encoding vectors satisfy, 
Note that position encoding vectors $\mathbf{E}_{pos}^{(n)}$ are generated independently from $\mathcal{N}(0,10)$, hence for $n\neq t$,
\vspace{-0.1cm}\begin{equation} \label{eq:larger}
(\mathbf{E}_{pos}^{(t)})^{\text{T}}\mathbf{E}_{pos}^{(t)}>> (\mathbf{E}_{pos}^{(t)})^{\text{T}}\mathbf{E}_{pos}^{(n)}
\vspace{-0.1cm}\end{equation} 
As a result, the attention matrix becomes an identity matrix,
\begin{equation}
    \mathbf{A}^h \triangleq softmax((\mathbf{Q}^h)^{\text{T}}\mathbf{K}^h/\sqrt{D_h}) 
    \cong \mathbf{I}_{(N+1) \times (N+1)} \notag
\end{equation}
After multiplying with the value matrix $\mathbf{V}^h$, the self-attention output for head $h$ is  $SA_h([\mathbf{y}^{(0,m)} \cdots \mathbf{y}^{(N,m)}]) \triangleq \mathbf{A}^h(\mathbf{V}^h)^{\text{T}}$.  
Let $\mathbf{W}_{MSA} \in \mathbb{R}^{D \times D}$ denote the  weight matrix that performs the  linear transformation on the concatenated outputs from all heads. 
We set $\mathbf{W}_{MSA} = \mathbf{I}_{D \times D}$, 
after which the MSA output is, 
\vspace{-0.1cm}\begin{align}
    &MSA([\mathbf{y}^{(0,m)} \hspace{0.2cm}\cdots \hspace{0.2cm} \mathbf{y}^{(N,m)}]) \notag \\
    &\triangleq \big[SA_1([\mathbf{y}^{(0,m)}  \hspace{0.2cm} \cdots  \hspace{0.2cm} \mathbf{y}^{(N,m)}]) \hspace{0.5cm}\cdots \notag\\
    &\hspace{1cm}  SA_L([\mathbf{y}^{(0,m)} \hspace{0.2cm} \cdots  \hspace{0.2cm} \mathbf{y}^{(N,m)}])\big] \times \mathbf{W}_{MSA} \\
    &\cong \begin{bmatrix}
         (\mathbf{y}^{(0,m)}) & 
        \cdots&
        (\mathbf{y}^{(N,m)})
    \end{bmatrix}_{(N+1) \times D}^{\text{T}} \label{eq:MSAop}
\end{align}
which ensures the flow of $\mathbf{y}^{(n,m)}$ for $n\in[N]$, $m\in[M]$.  

\begin{figure}
  \centering
   \includegraphics[width=0.8\linewidth]{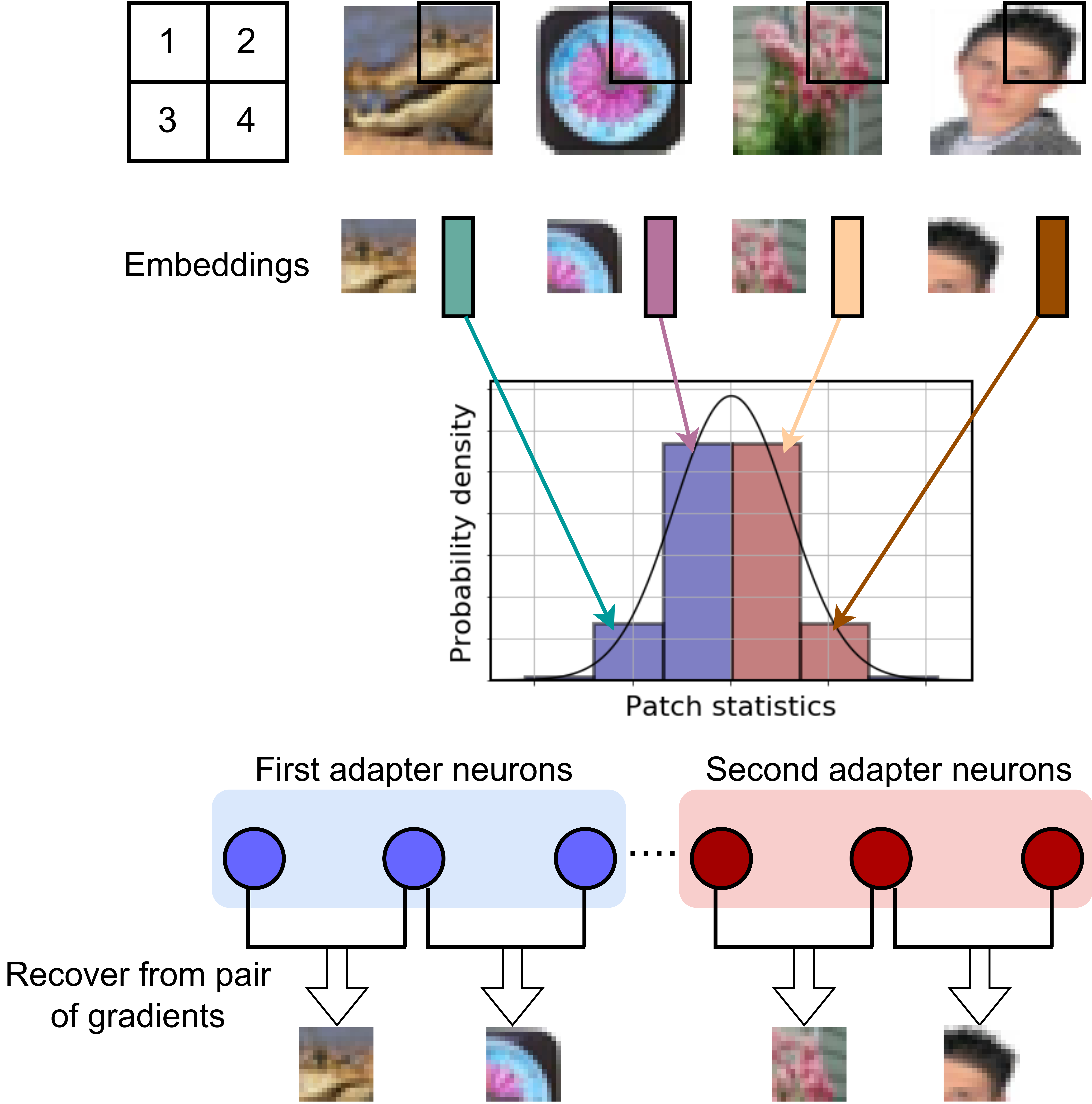}
   %\vspace{-1cm}
   \caption{{\bf Image recovery from multiple adapters.} A small-scale example with two adapters, each consisting of $3$ neurons in the down-projection layer. The attacker aims to recover $4$ images in the batch, where each image is divided into $4$ patches. The biases for the two adapters are designed to recover patches from the second position. The plot represents the distribution of patch statistics, $(\mathbf{E}_{pos}^{(t)})^{\text{T}} \mathbf{x}_{map}^{(t,m)}$ for $t=2$. The attacker targets images whose patch statistics lie in one of the four intervals (colored). Each adapter targets two intervals. By leveraging multiple adapters, the attacker can recover all the patches in the target position. } 
   \label{fig:adapter}
\vspace{-0.4cm}\end{figure}

\subsection{Adapter Layer} \label{sec:adapt}

%\noindent
%{\bf Adapter. } 
The next layer is the adapter, consisting of two linear layers (one down-projection and one up-projection) with an activation function in-between. Down-projection projects the input embeddings of dimension $D$ to a lower dimension $r<<D$. 
Recall that adapter modules are trainable. 
% , while all other layers are frozen. 
By using the gradients in the down-projection layer along with the activation, one can try to recover the input embeddings $\{\mathbf{y}^{(n,m)}\}_{n \in [N], m \in [M]}$ as in \eqref{eq:base_attack}.   
On the other hand, the number of neurons $r$ typically ranges from $1$ to $64$ \cite{Houlsby2019, Shoufa2022}, which severely limits the number of images recovered, as reconstruction performance in  \eqref{eq:base_attack} degrades when the number of neurons decreases  \cite{Fowl2022, LOKI2024}. 
% leading to a limited set of weight and bias gradients 
To enable recovery of a large batch of images, we propose to use the neurons from multiple adapters, where different layers are designed to recover images with different statistical properties.

We first discuss how to recover a patch from a target position within an image. For a target position $t\in[N]$, the goal is to allow the patches from position $t$ pass the activation function, and filter out the patches from all other positions $n \neq t$. We allocate $k_t>r$ neurons to recover patches from  position $t$. As each adapter has $r$ neurons, $S_t \triangleq \frac{k_t}{r}$ adapters are used to recover patches from position $t$, and $S \triangleq \sum_{n=1}^{N}S_n$ adapters to recover the patches from all positions $t\in[N]$. 
Denote the set of neurons used for position $t$ as $\mathcal{N}_t$.  
 For each neuron $j \in \mathcal{N}_t$, we set the  corresponding weight vector to $\mathbf{E}_{pos}^{(t)}$, after which the output for patch $t$ is,  
\vspace{-0.1cm}\begin{align} 
    v_j^{(t,m)}&= (\mathbf{E}_{pos}^{(t)})^{\text{T}} \mathbf{y}^{(t,m)}+b_j\\
    &=(\mathbf{E}_{pos}^{(t)})^{\text{T}} \mathbf{x}_{map}^{(t,m)}+(\mathbf{E}_{pos}^{(t)})^{\text{T}}\mathbf{E}_{pos}^{(t)}+b_j \label{eq:neuron_target}
\end{align}
whereas for all other positions $n \in \{0, \ldots, N\} \backslash \{t\}$,
\vspace{-0.1cm}\begin{align}
    v_j^{(n,m)}&= (\mathbf{E}_{pos}^{(t)})^{\text{T}} \mathbf{y}^{(n,m)}+b_j \notag \\
    &=(\mathbf{E}_{pos}^{(t)})^{\text{T}} \mathbf{x}_{map}^{(n,m)}+(\mathbf{E}_{pos}^{(t)})^{\text{T}}\mathbf{E}_{pos}^{(n)}+b_j 
\end{align}

We next discuss  how this design plays a key role in propagating patches from target position $t$ while blocking patches from all other positions. Note that the server does not know  $(\mathbf{E}_{pos}^{(t)})^{\text{T}}\mathbf{x}_{map}^{(t,m)}$ as it does not have access to local data. Along the lines of \cite{Fowl2022, Fowl2023}, we assume the server can instead estimate an approximate distribution for this quantity by utilizing a public dataset, which resembles a Gaussian distribution from the central limit theorem \cite{Fowl2022}. 
Define, 
\begin{align} \label{eq:CDF}
   c_j \triangleq \psi^{-1}(j/k_t)
\end{align}
where $\psi^{-1}(\cdot)$ is the inverse CDF of the estimated Gaussian. 
Then, the bias for neuron $j$ is designed as,
\begin{align} \label{eq:design_bias2}
    b_j \triangleq -(\mathbf{E}_{pos}^{(t)})^{\text{T}}\mathbf{E}_{pos}^{(t)} -c_j
\end{align} 
If for any given patch $t$ and image $m$,
\begin{equation} \label{eq:condn3}
  c_j<  (\mathbf{E}_{pos}^{(t)})^{\text{T}} \mathbf{x}_{map}^{(t,m)}<c_{j+1}
\end{equation}
then,  from \eqref{eq:neuron_target}, \eqref{eq:design_bias2} and \eqref{eq:condn3} we observe for patch $t$, 
\begin{equation}
    v_j^{(t,m)}= (\mathbf{E}_{pos}^{(t)})^{\text{T}} \mathbf{y}^{(t,m)}+b_j
    =(\mathbf{E}_{pos}^{(t)})^{\text{T}} \mathbf{x}_{map}^{(t,m)}-c_j>0 \label{eq:neuron_j}
\end{equation}
whereas for all other patches $n \in \{0, \ldots, N\} \backslash \{t\}$,
\begin{align}
    v_j^{(n,m)}&= (\mathbf{E}_{pos}^{(t)})^{\text{T}} \mathbf{y}^{(n,m)}+b_j <<0 
    \label{eq:negative}
\end{align}
which follows from \eqref{eq:larger}, hence \eqref{eq:negative} filters out the patches from all other positions after the activation function.   
Next, for the target position $t$, we need 
to ensure unique recovery of each image patch in the batch. For this, we utilize the technique from \cite{Fowl2022} to successively block images by the activation function. According to \eqref{eq:neuron_target}, \eqref{eq:design_bias2} and \eqref{eq:condn3}, 
\begin{align}
    v_{j+1}^{(t,m)}  
    % = (\mathbf{E}_{pos}^{(t)})^{\text{T}} \mathbf{y}^{(t,m)}+b_{j+1}  
    = (\mathbf{E}_{pos}^{(t)})^{\text{T}} \mathbf{x}_{map}^{(t,m)}-c_{j+1}<0 \label{eq:neuron_{j+1}} 
\end{align}
hence $\mathbf{y}^{(t,m)}$ propagates through the activation function at neuron $j$ as in \eqref{eq:neuron_j}, but is blocked at neuron $j+1$ as in \eqref{eq:neuron_{j+1}}.  If no other image satisfies \eqref{eq:condn3}, then by leveraging the gradients for this pair of neurons, the attacker can recover,
\begin{align} \label{eq:RobFed}
  \bigg (\frac{\partial \mathcal{L}_i} {\partial \mathbf{w}_j}-\frac{\partial \mathcal{L}_i} {\partial \mathbf{w}_{j+1}} \bigg )\bigg / \bigg ( \frac{\partial \mathcal{L}_i} {\partial b_j}-\frac{\partial \mathcal{L}_i} {\partial b_{j+1}} \bigg ) = \mathbf{y}^{(t,m)}
\end{align}
After computing $c_j$ for $j \in [k_t]$ as in \eqref{eq:CDF}, we can utilize $c_{(s-1)r+1}, \ldots, c_{sr}$ to design the biases of $r$ neurons as in \eqref{eq:design_bias2} for the  $s^{th}$ adapter for $s \in [S_t]$. 
As long as only one patch lies within an interval $[c_j, c_{j+1}]$ for $j \in [k_t]$, perfect recovery can be achieved for that particular patch. From a single adapter, the attacker can recover at most $r-1$ patches that lie within one of the $r-1$ intervals. Attacks to more intervals can be deployed by utilizing $S_t$ adapters for position $t$. We demonstrate an illustrative example in Fig.~\ref{fig:adapter}. 
After recovering the embedding $\mathbf{y}^{(t,m)}$, the attacker can recover, 
\begin{align} \label{eq:final_patch}
    \mathbf{x}^{(t,m)}= \mathbf{E}^{\dagger}(\mathbf{y}^{(t,m)}-\mathbf{E}_{pos}^{(t)})
\end{align}
where $\mathbf{E}^{\dagger}$ is the pseudoinverse of $\mathbf{E}$ in \eqref{eq:x_map}.
The weight and bias for the up-projection are designed to produce zero output, to ensure we regain the target  embeddings $\{\mathbf{y}^{(n,m)}\}_{n \in [N], m \in [M]}$ after the residual connection in Fig.~\ref{fig:ViT}. 
However, if we simply set all parameters to zero,  the gradient for the preceding layer will be zero during backpropagation, making reconstruction impossible. We prevent this by setting the first neuron's weight   to a small non-zero value ($\sim10^{-6}$). We next discuss how to propagate the current layer's output (embeddings $\mathbf{y}^{(n,m)}$), from one adapter layer to the next to continue  reconstruction.

\begin{figure*}
  \centering 
   \begin{subfigure}{0.45\linewidth}
    \includegraphics[width=\linewidth]{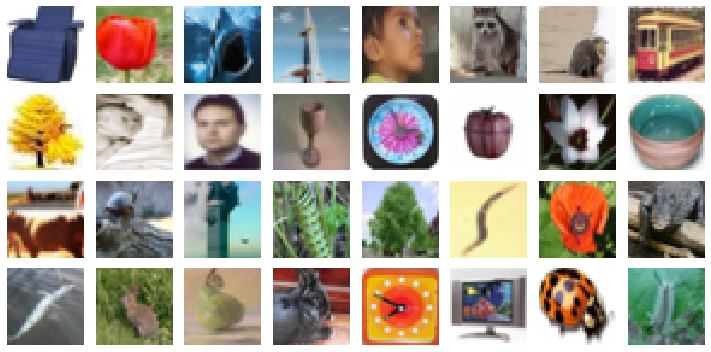} 
    \caption{Original images}
    \label{fig:original_CIFAR100}
  \end{subfigure}
  \hspace{0.5cm}
  \begin{subfigure}{0.45\linewidth}
    \includegraphics[width=\linewidth]{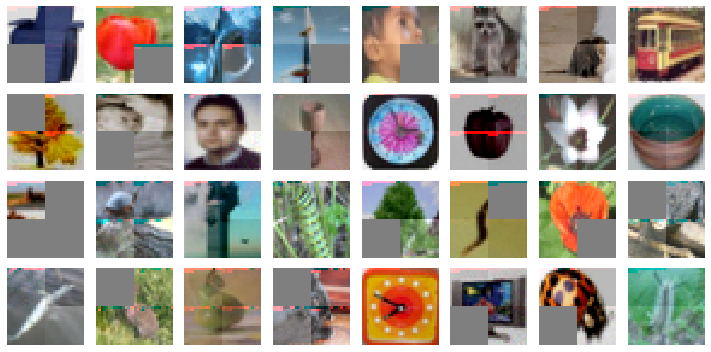}  
    \caption{Recovered}
    \label{fig:rec_CIFAR100}
  \end{subfigure}
\vspace{-0.2cm}  \caption{CIFAR-100 (recovered images for a batch of $32$ images). }
  \label{fig:recovery_CIFAR100}
\vspace{-0.3cm}\end{figure*}

%\noindent
%{\bf Second LayerNorm (LN2). } 

\subsection{Second LayerNorm (LN2)} \label{sec:LN2}

As shown in Fig. \ref{fig:ViT}, the adapter output goes through a residual connection from the LN1 input, which is equal to $\mathbf{y}^{(n,m)}$ from \eqref{eq:LN1}. 
The output of the residual connection is then,
\begin{align} \label{eq:residual2}
    \mathbf{e}^{(n,m)} \triangleq (1+1)\mathbf{y}^{(n,m)}  \cong 2\mathbf{y}^{(n,m)}
\end{align}
for $n \in \{0, \ldots, N\}$, which enters another LayerNorm (LN2). 
Note that $\mathbf{y}^{(n,m)}$ has mean approximately equal to $0$ and standard deviation $\sigma$ from \eqref{eq:LN1}. Therefore, $ \mathbf{e}^{(n,m)}$ has mean $0$ and standard deviation $2\sigma$. The server can then set the elements in the LN2 weight vector to $\sigma$ and bias vector to $0$ to retrieve back the  original embeddings. 
\begin{align} \label{eq:LN2}
     \frac{\mathbf{e}^{(n,m)}-0}{2\sigma}\sigma +0 \cong \mathbf{y}^{(n,m)}
\end{align}

%\noindent
%{\bf Multi Layer Perceptron (MLP) Layer. } 

\subsection{Multi Layer Perceptron (MLP)} \label{sec:MLP}

Next is the MLP layer, consisting of two linear layers with a GELU activation in-between. 
The first layer maps the embeddings to a higher dimension $4D$, whereas the second layer maps them back to the original dimension $D$ \cite{Dosovitskiy2021}. Our goal is to propagate the embeddings $\mathbf{y}^{(n,m)}$ to the next adapter undistorted. For this, the MLP needs to act as an identity function. Let $\mathbf{W}_{MLP,1} \in \mathbb{R}^{4D \times D}$ be the weight matrix  in the first layer,  with row $p$ and column $q$ given as,
\begin{align} \label{eq:MLP1_weight}
\mathbf{W}_{MLP,1}[p,q] \triangleq \left\{ \begin{matrix} 1 &\text{if  }p=q\\
0 &\text{ otherwise}
\end{matrix} \right.
\end{align}
which makes the output equal to the original embeddings, $\mathbf{y}^{(n,m)}$ in the first $D$ coordinates, while the remaining $3D$ coordinates are $0$. Note that original embeddings may contain negative values with a large magnitude, which will be filtered out by GELU, whereas values close to zero are subject to attenuation. To avoid this, we set the bias as $\mathbf{b}_{MLP,1}  \triangleq \gamma \mathbf{1}_{4D}$ with a large value $\gamma = 10^4$, where $\mathbf{1}_{4D}$ is a $4D$-dimensional vector containing all $1$s. 
 This ensures that the neuron outputs are not affected by GELU. By setting the weight matrix $\mathbf{W}_{MLP,2} \in \mathbb{R}^{D \times 4D}$ and bias $\mathbf{b}_{MLP,2} \in \mathbb{R}^D$ in the second linear layer as,
\vspace{-0.12cm}\begin{equation} \label{eq:MLP2_weight}
    \mathbf{W}_{MLP,2}[p,q] \triangleq \left\{ \begin{matrix} 1 &\text{ if } p=q\\
    0 &\text{ otherwise } \end{matrix} \right. 
\vspace{-0.12cm}\end{equation}
and $\mathbf{b}_{MLP,2} \triangleq  -\gamma \mathbf{1}_{D}$, we ensure that the MLP output is equal to the original embeddings $\mathbf{y}^{(n,m)}$ for $n \in \{0, \ldots , N\}$, $m \in [M]$. 
\begin{figure*}
  \centering 
   \begin{subfigure}{0.45\linewidth}
    \includegraphics[width=\linewidth]{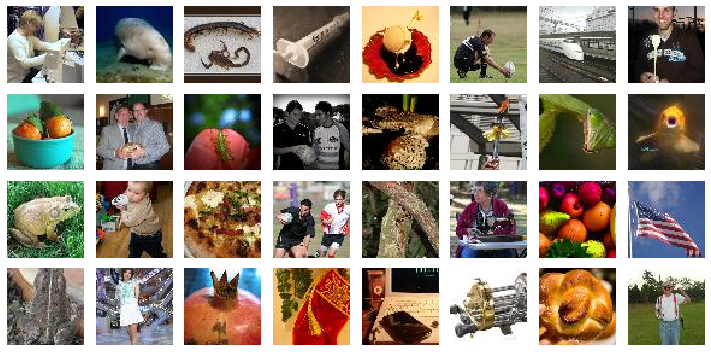}  
    \caption{Original images}
    \label{fig:original_TinyImageNet}
  \end{subfigure}
  \hspace{0.5cm}
  \begin{subfigure}{0.45\linewidth}
    \includegraphics[width=\linewidth]{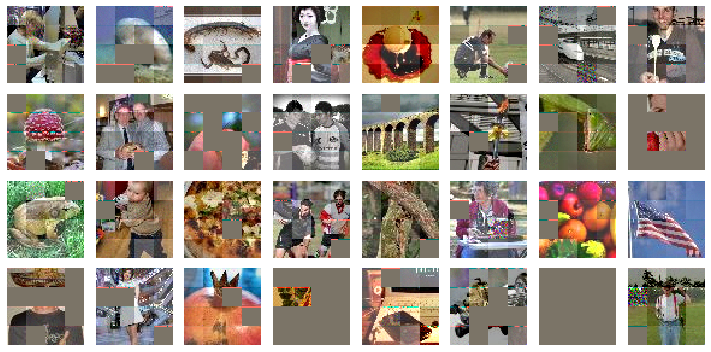}  
    \caption{Recovered}
    \label{fig:rec_TinyImageNet}
  \end{subfigure}
\vspace{-0.1cm}  \caption{TinyImageNet (recovered images for a batch of $32$ images). }
  \label{fig:recovery_TinyImageNet}
\vspace{-0.4cm}\end{figure*} 
We adopt the same design  in the subsequent LayerNorm, MSA and MLP layers. This ensures that the target embeddings $\mathbf{y}^{(n,m)}$ propagate through all intermediate layers towards the adapters. 
After recovering raw image patches, the final step is to group the patches that belong to the same image. 
For this, we use a similar approach to \cite{Fowl2023}, which embeds a unique tag in the first MSA layer to each patch belonging to the same source image.

\section{Experiments} \label{sec:exp}
\begin{table}[t]
\centering
 % \small
\footnotesize

 % \addtolength{\tabcolsep}{-0pt} 
\setlength{\tabcolsep}{3pt}
\begin{tabular}{@{}lc c c c  c c @{}}\toprule

& \multicolumn{2}{c}{\bf LPIPS} & \multicolumn{2}{c}{\bf SSIM}   & \multicolumn{2}{c}{\bf MSE} \\
% \cmidrule(lr){2-4} \cmidrule(lr){5-7}
% \midrule 

%& avg & min &max & avg & min &max & avg & min &max & avg & min &max\\ 
 % \midrule

 %\cmidrule(lr){8-10} \cmidrule(lr){11-13}
% \\\cmidrule(lr){2}
 &Mean &Std &Mean &Std &Mean &Std\\
  \cmidrule(lr){2-3}  \cmidrule(lr){4-5} \cmidrule(lr){6-7} 
%            & MSE   \\\midrule
CIFAR-10 & $0.10$ &$0.09$  &$0.74$ &$0.11$ &$0.21$ &$0.12$  \\
CIFAR-100 & $0.08$ &$0.06$  &$0.88$ &$0.12$ &$0.20$ &$0.16$ \\
TinyImageNet & $0.12$ &$0.04$  &$0.76$ &$0.10$ &$1.06$ &$0.82$ \\
   \bottomrule
\end{tabular} 
% \addtolength{\tabcolsep}{2pt}

  % \vspace{-0.1cm}
  \caption{Mean and standard deviation of MSE, LPIPS and SSIM scores for different datasets. }
  %Lower LPIPS/higher SSIM implies better reconstruction.  } 
\label{table:metric}
 \vspace{-0.3cm}
\end{table}

\begin{figure}
  \centering
  % Subfigure 1
   \begin{subfigure}{0.45\linewidth}
    \includegraphics[width=\linewidth]{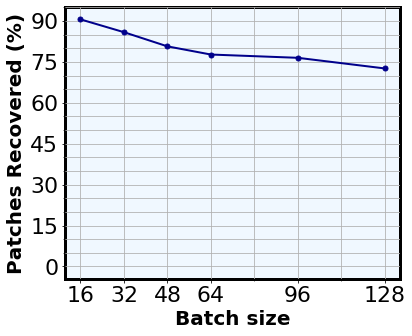}  % Insert your first image here
    \caption{Varying batch size}
    \label{fig:CIFAR100_batch}
  \end{subfigure}
  %\hspace{0.5cm}
  \begin{subfigure}{0.45\linewidth}
    \includegraphics[width=\linewidth]{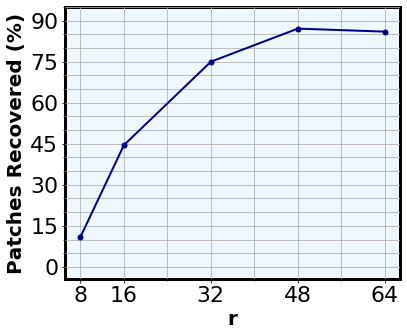}  
    \caption{Varying $r$}
    \label{fig:CIFAR100_r}
  \end{subfigure}\\
  \begin{subfigure}{0.45\linewidth}
    \includegraphics[width=\linewidth]{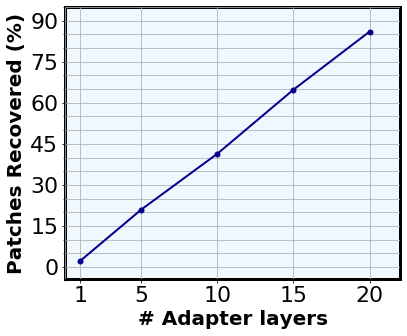}  
    \caption{Varying $\#$ adapter layers}
    \label{fig:CIFAR100_layer}
  \end{subfigure}
   \vspace{-0.15cm} \caption{Percentage of patches  recovered with varying batch size, bottleneck dimension, and number of adapter layers used. }
  \label{fig:plots}
   \vspace{-0.1cm}\end{figure}

\begin{figure}
  \centering
  % Subfigure 1
  %  \begin{subfigure}{0.305\linewidth}
  %   \includegraphics[width=\linewidth]{figures/vary_round_GT.png}  % 
  %   \caption{Ground-truth}
  %   \label{fig:GT}
  % \end{subfigure}
   \begin{subfigure}{0.21\linewidth}
    \includegraphics[width=\linewidth]{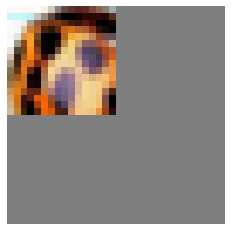}  % 
    \caption{Round 1}
    \label{fig:round1}
  \end{subfigure}
  %\hspace{0.5cm}
  \begin{subfigure}{0.21\linewidth}
    \includegraphics[width=\linewidth]{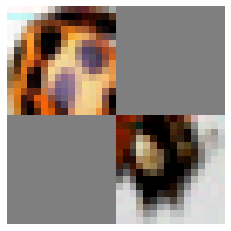}  
    \caption{Round 2}
    \label{fig:round2}
  \end{subfigure}
  \begin{subfigure}{0.21\linewidth}
    \includegraphics[width=\linewidth]{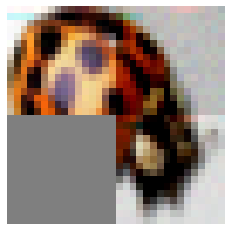}  
    \caption{Round 4 }
    \label{fig:round4}
  \end{subfigure}
\begin{subfigure}{0.21\linewidth}
    \includegraphics[width=\linewidth]{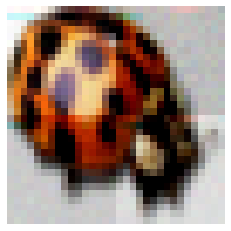}  
    \caption{Round 5 }
    \label{fig:round5}
  \end{subfigure}
  \vspace{-0.15cm}\caption{Patches recovered over multiple rounds ($r=8$). }
  \label{fig:multi-round}
\vspace{-0.5cm}\end{figure}

Our experiments aim to answer the following questions:

\begin{itemize} \itemsep0.1em
\item How does \namespace perform in terms of reconstruction?

\item How is performance impacted with varying bottleneck dimension  $r$ inside the adapters?

\item How is performance impacted with increasing batch size?

\item How do we benefit from leveraging the layers from all the adapter modules  for reconstruction?

\end{itemize}
{\bf Setup.} We consider a distributed setting for image classification  using ViT-B/16  \cite{Dosovitskiy2021} as the pretrained model. Each user holds data samples from CIFAR-10, CIFAR-100 \cite{krizhevsky2009learning} and TinyImageNet \cite{Tinyimagenet} datasets. 
 The experiments are run on a 24 core AMD Ryzen device with NVIDIA RTX4000.

\noindent
{\bf Hyperparameters. } In our experiments, each image is divided into patches of size $(16, 16)$ in accordance with \cite{Dosovitskiy2021}. For CIFAR-10 and CIFAR-100, each image (with resolution $(32, 32)$) is divided into $4$ patches. For TinyImageNet, each image (with resolution $(64, 64)$) is divided into $16$ patches. The embedding dimension is $D=768$ \cite{Dosovitskiy2021}.  
The bottleneck dimension for the adapters is $r=64$. 

\noindent
{\bf Performance Metrics.} Attack performance is evaluated using the mean squared error (MSE), and  perceptual/structural similarity scores LPIPS \cite{Zhang2018} and SSIM \cite{SSIM2004} between recovered and ground-truth images. Lower MSE/LPIPS or higher SSIM implies better reconstruction.
 
\noindent
{\bf Results. } In Fig. \ref{fig:recovery_CIFAR100}, we present the reconstructed images from the local gradient of a target user. The gradient is derived from training on a batch of $32$ images from CIFAR-100. Since each image is divided into $4$ patches, there are $32 \times 4=128$ target patches that the attacker aims to recover. As we observe, $110$-out-of-$128$ image patches ($85.9\%$ of the patches) are recovered. Gray patches indicate that the corresponding ground-truth patches are not recovered. 
In Fig. \ref{fig:recovery_TinyImageNet}, we present the reconstructed images for TinyImageNet. Since in this case each image is divided into $16$ patches, the attacker aims for $32 \times 16=512$ different patches. We observe that around $81\%$ of the total patches are recovered. 
% , again implying the efficacy of the attack.
Reconstruction for CIFAR-10 is presented in App.~\ref{app:CIFAR10}.
In Table \ref{table:metric},  we  report the mean and standard deviation of MSE, LPIPS, and SSIM scores across the recovered patches.

We next study the impact of batch size, bottleneck dimension, and number of adapter layers on reconstruction.  These experiments are conducted using CIFAR-100. 
% For these, we consider  CIFAR-100. 

% In the following experiments, we focus on CIFAR-100.
\noindent
{\bf Reconstruction rate vs. batch size.} We first evaluate the impact of batch size on the success rate. Since reconstruction is performed patch-wise, we compare  the percentage of patches  recovered across different batch sizes. As we observe in Fig. \ref{fig:CIFAR100_batch}, even for a batch size as large as $128$, \namespace recovers up to $72.6\%$ of the patches.

\noindent
{\bf Reconstruction rate vs. bottleneck dimension.} We present the impact of bottleneck dimension $r$ on the attack efficiency in Fig. \ref{fig:CIFAR100_r}. As $r$ increases, more patches are retrieved. Higher values of $r$ imply more neurons in the adapter layer that can be used for recovery. This suggests that choosing a small $r$ could be a potential defense against privacy attacks. However, as we show later, this can be countered by using multiple training rounds.

\noindent
{\bf Reconstruction rate vs. number of adapter layers.} We further analyze the impact of the number of adapter layers on the attack success in Fig. \ref{fig:CIFAR100_layer}. We observe that as we use more layers for the attack, more patches can be recovered. Even though a single adapter may have insufficient neurons due to down-projection, the leakage rate is enhanced by leveraging multiple adapter layers. 

\noindent
{\bf Attack over multiple training rounds.} We next demonstrate how the attack can be executed over multiple training rounds to increase the number of reconstructed patches for a small bottleneck dimension, in particular for $r=8$. As mentioned earlier, each user sends the adapter gradients to the server in each training round, while the pretrained model is kept frozen. The server sends the aggregated adapter parameters in each training round back to the users. Hence, the server can tamper with these adapter parameters so that in each round, different sets of intervals from \eqref{eq:condn3} can be targeted. For the image in Fig. \ref{fig:multi-round}, we observe that only a single patch is recovered from gradients within a single training round. However, from gradients over $5$ training rounds, the entire image is recovered. Hence, choosing a small value for $r$ does not guarantee privacy. 
\begin{figure}
  \centering
   \includegraphics[width=0.7\linewidth]{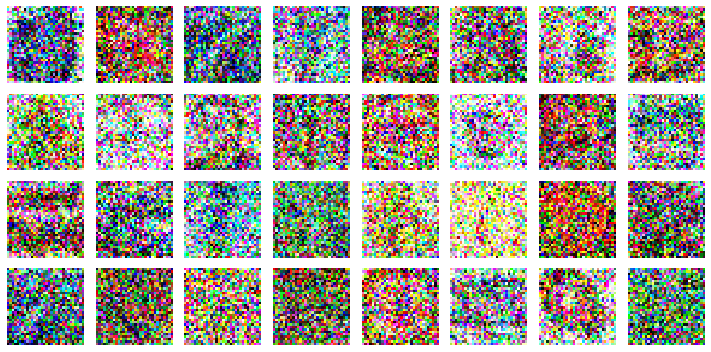}
   %\vspace{-1cm}
 \vspace{-0.05cm}  \caption{Recovered images with attack from \cite{Geiping2020} (CIFAR-100). } 
   \label{fig:CIFAR100_opt}
\vspace{-0.35cm}\end{figure}

\noindent
{\bf Comparison with  optimization-based baseline.} In Fig. \ref{fig:CIFAR100_opt}, we  apply the  optimization-based attack from \cite{Geiping2020} for the images in Fig. \ref{fig:original_CIFAR100} (details are provided in App~\ref{app:op}). 
\begin{table}[h]
\centering
 % \small
% \footnotesize
\scriptsize

 % \addtolength{\tabcolsep}{-0pt} 
\setlength{\tabcolsep}{3pt}
\begin{tabular}{@{}lc c c c c  @{}}\toprule

{\bf Batch size} &8 &16 & 32  &48 &64\\

  \cmidrule{2-6}
%            & MSE   \\\midrule
$\%$ {\bf Patches recovered} &$90$ & $84.7$ &$73.2$ &$66.5$ &$62.6$   \\
   \bottomrule
\end{tabular} 
% \addtolength{\tabcolsep}{2pt}

  % \vspace{-0.1cm}
 \vspace{-0.2cm}  \caption{Reconstruction with varying batch size (ImageNet). } 
\label{table:imagenet}
 \vspace{-0.2cm}
\end{table}

\noindent
{\bf Scalability to high resolution dataset.} Table \ref{table:imagenet} presents patch recovery  on ImageNet \cite{Deng2009}, with image resolution $(224, 224)$. Each image is divided into $196$ patches of size $(16,16)$.  
Additional experiments are provided in App.~\ref{app:exp}.

\section{Conclusion} \label{sec:conc}
We show how an adversarial server can manipulate the pretrained model and adapter parameters to uncover users' local fine-tuning data in PEFT for FL, despite a significantly reduced embedding space. 
Our attacks demonstrate the critical need for privacy-aware PEFT mechanisms, with future directions including certifiable privacy guarantees. 

\section{Acknowledgement}
The research was supported in part by the OUSD (R$\&$E)/RT$\&$L under
Cooperative Agreement W911NF-20-2-0267, NSF CAREER Award
CCF-2144927, ARO grant W911NF2210260, and the UCR OASIS Grant.  The views and conclusions contained in this document are those of the authors and should not be interpreted as representing the official policies, either expressed or implied, of the ARL and OUSD(R$\&$E)/RT$\&$L or the U.S. Government.

{
    \small
    \bibliographystyle{ieeenat_fullname}
    \bibliography{ref}
}

\clearpage

\noindent 
{\bf  \LARGE Appendix} 

%\afterpage{

\setcounter{section}{0}
% Change section numbering to use letters
\renewcommand{\thesection}{\Alph{section}}

%\section*{{\bf  \LARGE Appendix}}

\section{Ethical Considerations}
% \begingroup
% \renewcommand{\floatpagefraction}{0.8} % Allow floats on less-filled pages
% \renewcommand{\topfraction}{0.9}       % Relax placement rules slightly
% \renewcommand{\bottomfraction}{0.9}
%\vspace{0.5cm}

Our work points to potential privacy threats that may occur when parameter-efficient fine-tuning (PEFT) is applied under the federated learning (FL) setup. Since (to the best of our knowledge) privacy concerns under PEFT based FL applications are under-explored, our observations suggest an important challenge that local data can be revealed if no additional defense mechanism is applied. Users involved in training might be oblivious to these risks. As a malicious server can deploy such attacks by merely poisoning model parameters, it is crucial to explore robust verification algorithms to examine the authenticity of the models received from the server. Furthermore, defense strategies such as differential privacy under the PEFT setting can prevent the server from observing the true local gradients with a small impact on utility. 
We hope that our work will motivate new research directions towards certifiable privacy, integrity, and authenticity guarantees for PEFT mechanisms. 
 
\section{Algorithm}
We provide the pseudocode of our proposed attack, \name, in Algorithm 1.

\raggedbottom

%\begin{multicols}{2}
%\begin{minipage}{\linewidth}
\begin{algorithm}[h]%[t]
% \small
 \footnotesize
	\caption{\name}  \label{Alg} 
 
	 \KwInput{Pretrained model $\mathbf{w}_F$, adapter parameters $\mathbf{w}_A$,  
   adapter gradients $\frac{\partial \mathcal{L}_i}{\partial \mathbf{w}_A}$ of user $i$ (victim)}    
 \KwOutput{ Recovered patches $\mathbf{x}^{(t,m)}$ for $t \in [N], m \in [M]$ of user $i$, where $N$ is the total number of patches and $M$ is the number of images in the batch}
	
% \vspace{0cm}
 \tcp{\textbf{Server: Poisoning pretrained model}, $\mathbf{w}_F$}
 %\Indp
  \tcp{\textit{(Position encoding vectors)}}
Select $\mathbf{E}_{pos}^{(n)} \sim \mathcal{N}(0,\sigma)$
  for $n \in \{0, \ldots, N\}$ \\
   %\Indm
   
   %\Indp
\tcp{\textit{(Linear embedding matrix)}}
 Set $\mathbf{E}$ in \eqref{eq:x_map} to $0.5\mathbf{I}_D$\\
 %\Indm

 %\Indp
\tcp{ \textit{(MSA layer parameters)} }
 Set $\mathbf{W}_Q^h$, $\mathbf{W}_K^h$, $\mathbf{W}_V^h=\mathbf{I}_{D_h \times D_h}$ for head $h \in [L]$ \hfill $\triangleright$ Equation \eqref{eq:weight_MSA}\\
 Set $\mathbf{b}_Q^h$, $\mathbf{b}_K^h$, $\mathbf{b}_V^h=\mathbf{0}$ for head $h \in [L]$ \hfill $\triangleright$ Equation \eqref{eq:bias_MSA}\\
 Set $\mathbf{W}_{MSA}=\mathbf{I}_{D \times D}$ \hfill $\triangleright$ Section \ref{sec:MSA}\\
 %\Indm

  %\Indp
\tcp{\textit{(MLP layer parameters)} }
  Design weights $\mathbf{W}_{MLP,1}, \mathbf{W}_{MLP,2}$  according to \eqref{eq:MLP1_weight}, \eqref{eq:MLP2_weight}\\
 Design biases $\mathbf{b}_{MLP,1}=\gamma \mathbf{1}_{4D}$, $\mathbf{b}_{MLP,2}=-\gamma \mathbf{1}_{D}$ \hfill $\triangleright$ Section \ref{sec:MLP}\\
 %\Indm
 
%  %\STATE Design bias  \hfill Section \ref{sec:MLP}\\
%\Indp
  \tcp{\textit{(LN1 and LN2 layer parameters)}}
Set weights $\mathbf{w}_{LN1}$, $\mathbf{w}_{LN2}=\sigma \mathbf{1}_D$ \hfill $\triangleright$ Sections \ref{sec:LN1}, \ref{sec:LN2}\\
 Set biases $\mathbf{b}_{LN1}$, $\mathbf{b}_{LN2}$ to $\mathbf{0}_D$ \hfill $\triangleright$ Sections \ref{sec:LN1}, \ref{sec:LN2}\\
 %\Indm
 
Send $\mathbf{w}_F$ to the users \hfill $\triangleright$ sent once prior to training\\
 \tcp{\textbf{Server: Poisoning global adapter}, $\mathbf{w}_A$ }
 Set weights in down-projection layer to $\mathbf{E}_{pos}^{(t)}$ for target position $t \in [N]$ \hfill $\triangleright$ Section \ref{sec:adapt}
 
Design biases in down-projection layer according to \eqref{eq:design_bias2}

Set weights and biases in up-projection layer to $0$ \hfill $\triangleright$ Section \ref{sec:adapt}

 Send $\mathbf{w}_A$ to the users \hfill $\triangleright$ sent in each training round
 
\tcp{\textbf{User $i$: Local training}}

Compute loss $\mathcal{L}_i(\mathbf{w}_F, \mathbf{w}_A)$ for batch of images

Compute gradient $\frac{\partial \mathcal{L}_i}{\mathbf{w}_A}$

Send $\frac{\partial \mathcal{L}_i}{\mathbf{w}_A}$ to the server \hfill $\triangleright$ sent in each training round

\tcp{\textbf{Server: Reconstruction from gradients}}
%  %\STATE Receive adapter gradients from the user
 Recover embeddings $\mathbf{y}^{(t,m)}$ for $t \in [N], m \in [M]$ \hfill $\triangleright$ Equation \eqref{eq:RobFed}
 
 Recover patch $\mathbf{x}^{(t,m)}$ for $t \in [N], m \in [M]$ \hfill $\triangleright$ Equation \eqref{eq:final_patch}
 
Return $\mathbf{x}^{(t,m)}$ for $t \in [N], m \in [M]$ \hfill $\triangleright$ recovered patches

\end{algorithm} 
%\end{minipage}
%\end{multicols}
\section{Additional Experiments} \label{app:exp}
% \begingroup
% \renewcommand{\floatpagefraction}{0.8} % Relax float-only page requirements
% \renewcommand{\dblfloatpagefraction}{0.8}

%\endgroup

%\clearpage

In this section, we provide additional experimental results for our proposed framework. Unless stated otherwise, for all the experiments below, we use a batch size $32$, bottleneck dimension $r=64$ and ViT-B/16 architecture in accordance with the experiments in Section \ref{sec:exp}.

\begin{figure*}
  \centering
  % Subfigure 1
   \begin{subfigure}{0.46\linewidth}
    \includegraphics[width=\linewidth]{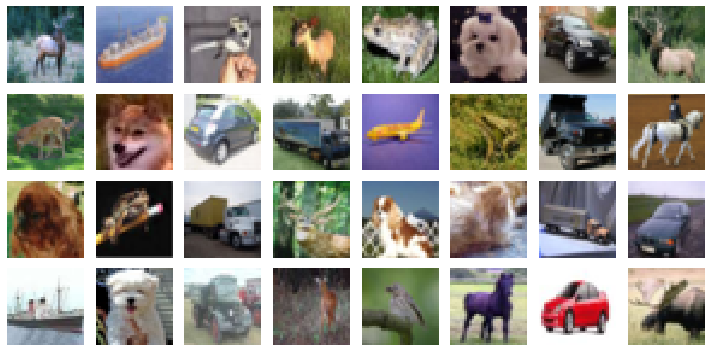}  
    \caption{Original images}
    \label{fig:original_CIFAR10}
  \end{subfigure}
  \hspace{0.5cm}
  \begin{subfigure}{0.46\linewidth}
    \includegraphics[width=\linewidth]{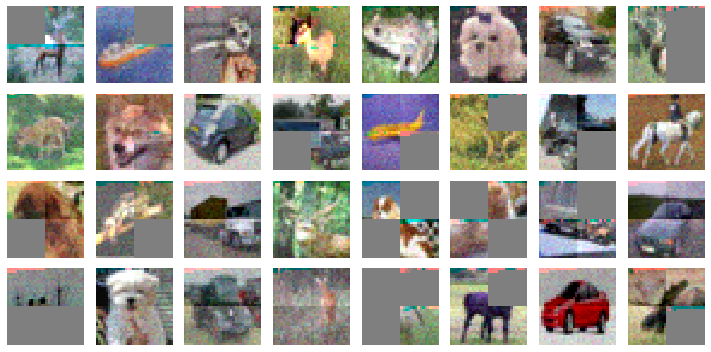}  
    \caption{Recovered}
    \label{fig:rec_CIFAR10}
  \end{subfigure}
 \vspace{-0.2cm} \caption{CIFAR-10 (recovered images for a batch of $32$ images). }
  \label{fig:recovery_CIFAR10}
\vspace{-0.1cm}\end{figure*}

\begin{figure*}[!htb]
  \centering
 \begin{subfigure}{0.46\linewidth}
    \includegraphics[width=\linewidth]{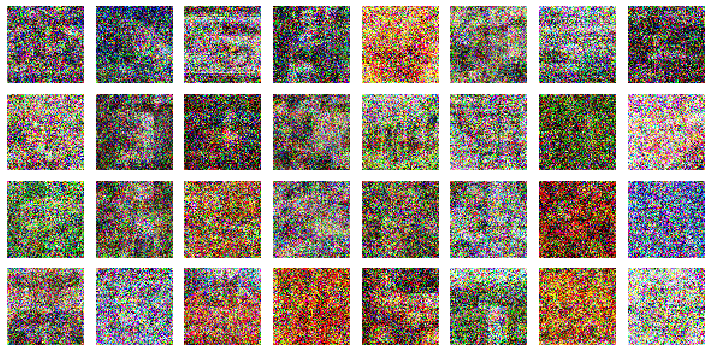}  
    \caption{Recovered from \cite{Geiping2020}}
    \label{fig:rec_opt_TinyImageNet}
  \end{subfigure}
   \hspace{0.5cm}
  \begin{subfigure}{0.46\linewidth}
    \includegraphics[width=\linewidth]{figures/TinyImageNet_final.png}  
    \caption{Recovered (\name)}
    \label{fig:rec_PEFTLeak_TinyImageNet}
  \end{subfigure}
  %\hspace{0.5cm}  
 \vspace{-0.2cm} \caption{Comparison with optimization-based benchmark from \cite{Geiping2020} (TinyImageNet). }
  \label{fig:recovery_opt_TinyImageNet}
\vspace{-0.1cm}\end{figure*}

\begin{figure*}[hbt!] %[t]
  \centering
 \begin{subfigure}{0.46\linewidth}
    \includegraphics[width=\linewidth]{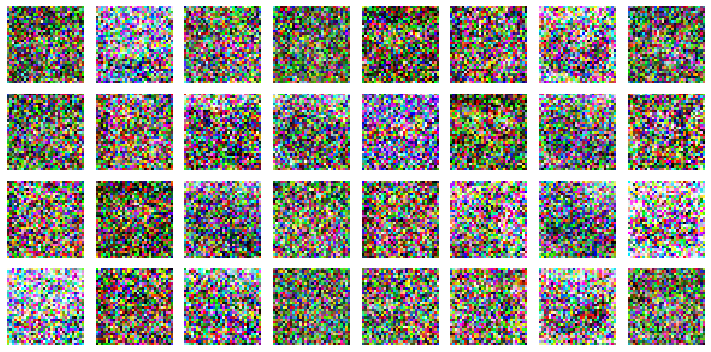}  
    \caption{Recovered from \cite{Geiping2020}}
    \label{fig:rec_opt_CIFAR10}
  \end{subfigure}
  \hspace{0.5cm}
  \begin{subfigure}{0.46\linewidth}
    \includegraphics[width=\linewidth]{figures/CIFAR10_final.png}  
    \caption{Recovered (\name)}
    \label{fig:rec_PEFTLeak_CIFAR10}
  \end{subfigure} 
 \vspace{-0.2cm} \caption{Comparison with optimization-based benchmark from \cite{Geiping2020} (CIFAR-10). }
  \label{fig:recovery_opt_CIFAR10}
\vspace{-0.1cm}\end{figure*} 

\subsection{Recovered Images for CIFAR-10} \label{app:CIFAR10}
In Fig.~\ref{fig:recovery_CIFAR10}, we demonstrate the recovery of a batch of $32$ images from the gradient for the CIFAR-10 dataset. As we observe, $106$-out-of-$128$ image patches, i.e., $82.8\%$ of the patches are recovered.

\subsection{Comparison with the Optimization-Based Baseline} \label{app:op}
%In this section, we describe the details of the optimization-based gradient inversion attack baseline from \cite{Geiping2020}.  
We now describe  the details of the optimization-based gradient inversion attack baseline. 
 To the best of our knowledge, there are no successful optimization-based attack baselines under the PEFT setting. Reference \cite{Zhang2023} studied the performance of the optimization-based attack from \cite{Zhu2019} for PEFT and observed that it was not successful under the PEFT setup.   
 Attack from \cite{Geiping2020} improves over  \cite{Zhu2019} in terms of the reconstruction performance by taking the direction of the gradient into consideration.  
 Essentially, the goal is to find a batch of images $\mathbf{X}$ that minimize the cosine distance between the true gradient and the predicted gradient,
 
%\vspace{-0.2cm}
\begin{align}
\mathbf{X}^*= \arg \min _{\mathbf{X} } \mathcal{F} (\mathbf{X})
\end{align}
%\vspace{-0.2cm}
such that, 
\begin{align} \label{eq:TV}
 \mathcal{F} (\mathbf{X}) &\triangleq  1-\frac{\left\langle \Delta \mathbf{g}, \Delta \mathbf{g}^{pred}\right\rangle}{\big\|\Delta \mathbf{g}\big\|\big\| \Delta \mathbf{g}^{pred}\big\|} + T V (\mathbf{X})
\end{align}
where $\Delta \mathbf{g}$ is the actual gradient received from the victim user, $\Delta \mathbf{g}^{pred}$ is the predicted gradient from training on dummy images.  The total variation regularization $\text{TV}(\cdot)$ is used as a standard image prior to ensure the smoothness of the recovered image.  We note that this attack considers adversaries with limited capability, who do not adopt any malicious tampering with the protocol, such as changing the model parameters or architecture. 

We applied this attack to our PEFT setting and studied how well this gradient matching algorithm performs by leveraging the adapter gradients only. For this, we run the experiments for the images in Figs. \ref{fig:original_TinyImageNet} (in our main paper) and \ref{fig:original_CIFAR10} from TinyImageNet and CIFAR-10 datasets (CIFAR-100 results were already provided in Fig. \ref{fig:CIFAR100_opt} in our main paper.)  We demonstrate our results in Figs. \ref{fig:recovery_opt_TinyImageNet} and \ref{fig:recovery_opt_CIFAR10}, where we present the images reconstructed by the optimization-based attack vs. \name.  As we observe from Figs. \ref{fig:recovery_opt_TinyImageNet} and \ref{fig:recovery_opt_CIFAR10}, the optimization-based attack fails to reconstruct any of the images in the batch whereas \namespace recovers most of the images with high fidelity.

\begin{figure*}
  \centering
  \begin{subfigure}{0.3\linewidth}
    \includegraphics[width=\linewidth]{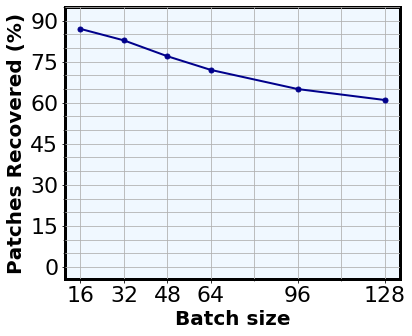}  % Insert your first image here
    \caption{Varying batch size}
    \label{fig:CIFAR10_Batch}
  \end{subfigure}
  \begin{subfigure}{0.3\linewidth}
    \includegraphics[width=\linewidth]{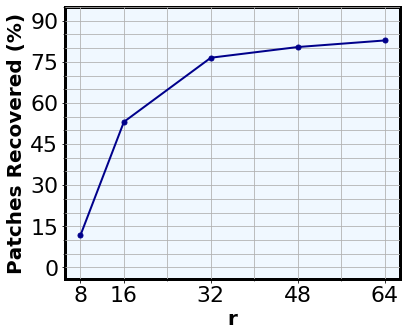}  
    \caption{Varying $r$}
    \label{fig:CIFAR10_r}
  \end{subfigure}
  \begin{subfigure}{0.3\linewidth}
    \includegraphics[width=\linewidth]{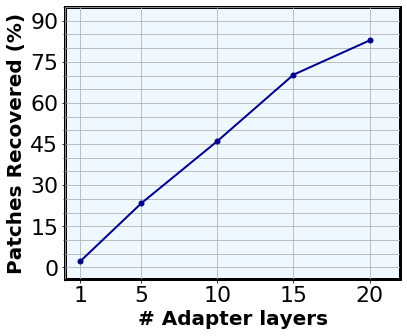}  
    \caption{Varying $\#$ adapter layers}
    \label{fig:CIFAR10_layer}
  \end{subfigure}
   \vspace{-0.05cm} \caption{Percentage of patches  recovered with varying batch size, bottleneck dimension, and number of adapter layers used within a single training round (CIFAR-10). }
  \label{fig:plots_CIFAR10}
\end{figure*}

\begin{figure*}
  \centering
  \begin{subfigure}{0.3\linewidth}
    \includegraphics[width=\linewidth]{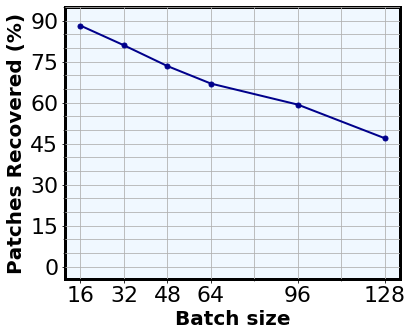}  % Insert your first image here
    \caption{Varying batch size}
    \label{fig:TinyImageNet_batch}
  \end{subfigure}
  \begin{subfigure}{0.3\linewidth}
    \includegraphics[width=\linewidth]{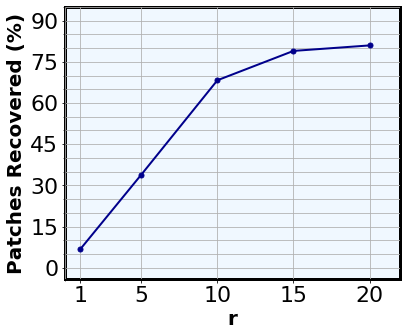}  
    \caption{Varying $r$}
    \label{fig:TinyImageNet_r}
  \end{subfigure}
  \begin{subfigure}{0.3\linewidth}
    \includegraphics[width=\linewidth]{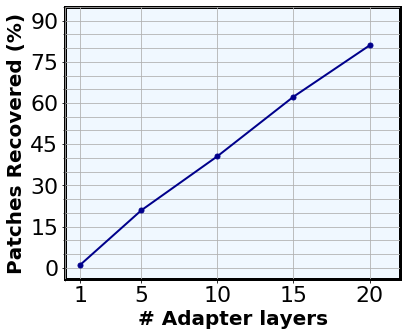}  
    \caption{Varying $\#$ adapter layers}
    \label{fig:TinyImageNet_layer}
  \end{subfigure}
   \vspace{-0.05cm} \caption{Percentage of patches  recovered with varying batch size, bottleneck dimension, and number of adapter layers used within a single training round (TinyImageNet). }
  \label{fig:plots_TinyImageNet}
\end{figure*}

\begin{table}[t]
\centering
 % \small
% \footnotesize
\scriptsize

 % \addtolength{\tabcolsep}{-0pt} 
\setlength{\tabcolsep}{3pt}
\begin{tabular}{@{}lc c c c  @{}}\toprule  
{\bf Architecture} &\shortstack{ViT-B/16 } & \shortstack{ViT-L/16 }  &\multicolumn{2}{c}{\shortstack{ViT-B/32}}\\
\midrule
% & &  &Naive &Improved \\
  % \cmidrule{4-5}
%            & MSE   \\\midrule
$\%$ {\bf Patches recovered} & $81$  &$81$ &$20.2$ (naive) &$79.6$ (improved) \\
   \bottomrule
\end{tabular} 
% \addtolength{\tabcolsep}{2pt}

  % \vspace{-0.1cm}
  \vspace{-0.15cm} \caption{Reconstruction for a batch of $32$ images  (TinyImageNet). } 
\label{table:arch}
 \vspace{-0.2cm}
\end{table}

\begin{figure}
  \centering
    \includegraphics[width=1\linewidth]{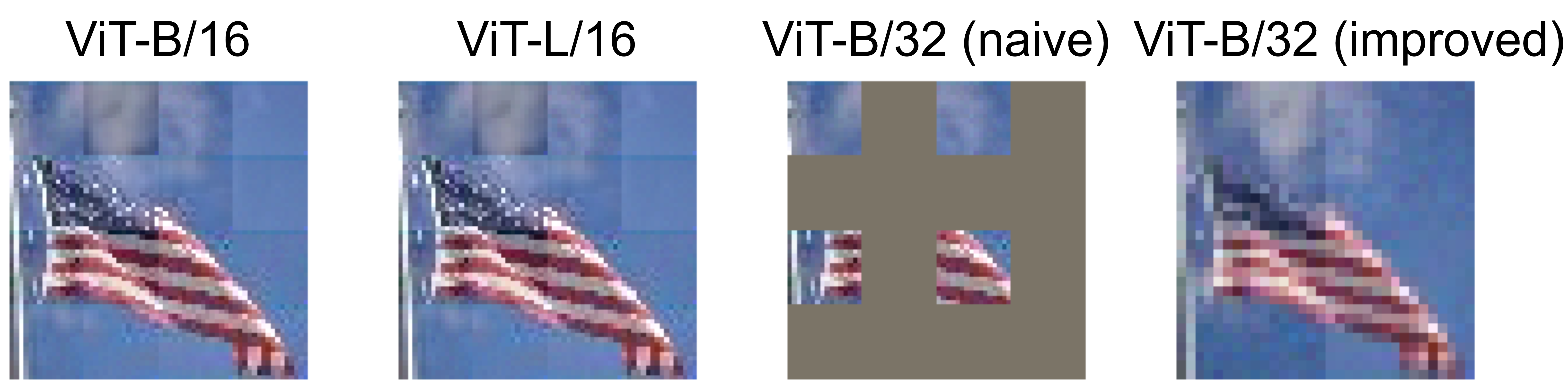} 
    %\caption{ViT-B/32 (improved)}
   % \label{fig:B_32_low}
 % \end{subfigure}
   \vspace{-0.3cm} \caption{Recovered images from different model architectures. }
  \label{fig:arch}
\vspace{-0.2cm} \end{figure}

\begin{figure*}
  \centering
  \begin{subfigure}{0.22\linewidth}
    \includegraphics[width=\linewidth]{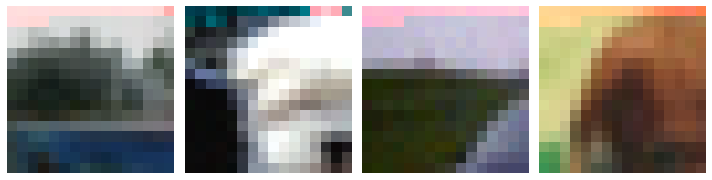}  
    \caption{$1^{\text{st}}$ layer}
    \label{fig:layer1_CIFAR10}
  \end{subfigure}
  \hspace{0.5cm}
  \begin{subfigure}{0.11\linewidth}
    \includegraphics[width=\linewidth]{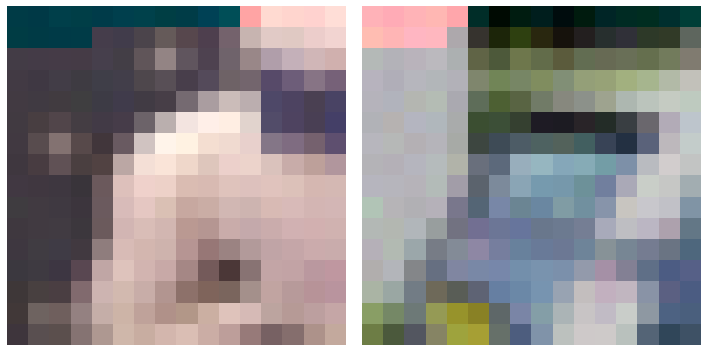}  
    \caption{ $2^{\text{nd}}$ layer}
    \label{fig:layer2_CIFAR10}
  \end{subfigure}
  \hspace{0.5cm}
  \begin{subfigure}{0.40\linewidth}
    \includegraphics[width=\linewidth]{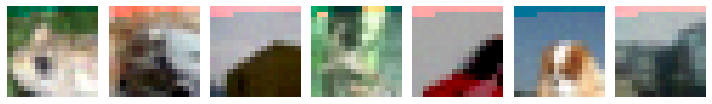}  
    \caption{$3^{\text{rd}}$ layer}
    \label{fig:layer3_CIFAR10}
  \end{subfigure}
  \hspace{0.5cm}
  \begin{subfigure}{0.29\linewidth}
    \includegraphics[width=\linewidth]{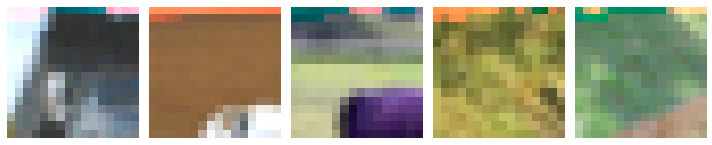}  
    \caption{$4^{\text{th}}$ layer}
    \label{fig:layer4_CIFAR10}
  \end{subfigure}
  \hspace{0.5cm}
  \begin{subfigure}{0.61\linewidth}
    \includegraphics[width=\linewidth]{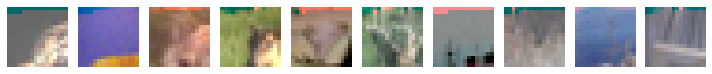}  
    \caption{$5^{\text{th}}$ layer}
    \label{fig:layer5_CIFAR10}
  \end{subfigure}
 \vspace{-0.2cm} \caption{Recovered patches from the first position using multiple adapter layers (CIFAR-10). }
  \label{fig:recovery_layer_CIFAR10}
\vspace{-0.1cm}\end{figure*}

\begin{figure*}
  \centering
  \begin{subfigure}{0.18\linewidth}
    \includegraphics[width=\linewidth]{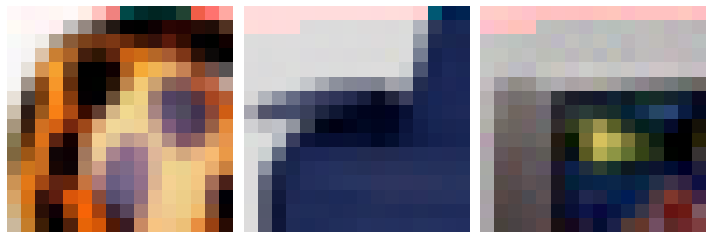}  
    \caption{$1^{\text{st}}$ layer}
    \label{fig:layer1}
  \end{subfigure}
  \hspace{0.5cm}
  \begin{subfigure}{0.42\linewidth}
    \includegraphics[width=\linewidth]{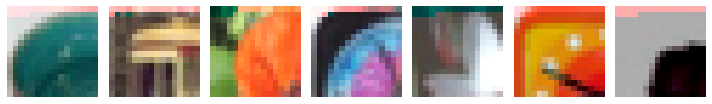}  
    \caption{$2^{\text{nd}}$ layer}
    \label{fig:layer2}
  \end{subfigure}
  \hspace{0.5cm}
  \begin{subfigure}{0.18\linewidth}
    \includegraphics[width=\linewidth]{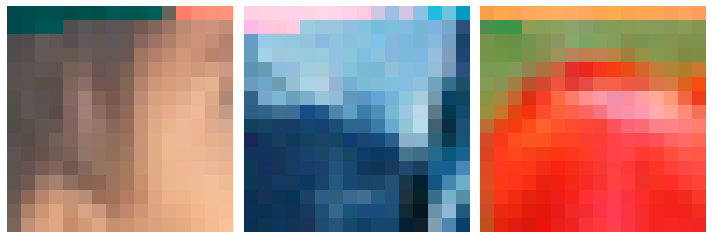}  
    \caption{$3^{\text{rd}}$ layer}
    \label{fig:layer3}
  \end{subfigure}
  \hspace{0.5cm}
  \begin{subfigure}{0.37\linewidth}
    \includegraphics[width=\linewidth]{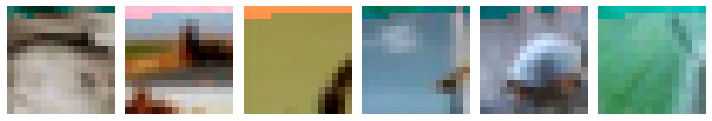}  
    \caption{$4^{\text{th}}$ layer}
    \label{fig:layer4}
  \end{subfigure}
  \hspace{0.5cm}
  \begin{subfigure}{0.58\linewidth}
    \includegraphics[width=\linewidth]{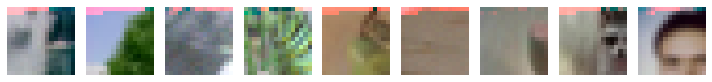}  
    \caption{$5^{\text{th}}$ layer}
    \label{fig:layer5}
  \end{subfigure}
 \vspace{-0.2cm} \caption{Recovered patches from the first position using multiple adapter layers (CIFAR-100). }
  \label{fig:recovery_layer_CIFAR100}
\vspace{-0.1cm}\end{figure*}

\begin{figure*}
  \centering
  \hspace{0.5cm}
  \begin{subfigure}{0.28\linewidth}
    \includegraphics[width=\linewidth]{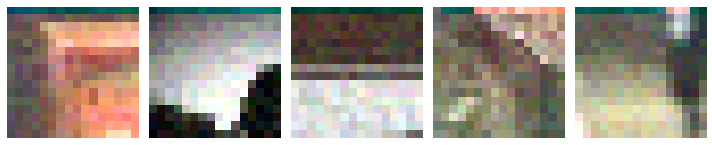}  
    \caption{$2^{\text{nd}}$ layer}
    \label{fig:layer2_Tinyimagenet}
  \end{subfigure}
  \hspace{0.5cm}
  \begin{subfigure}{0.7\linewidth}
    \includegraphics[width=\linewidth]{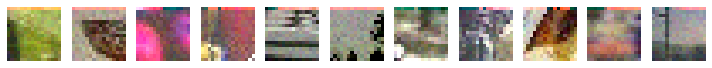}  
    \caption{$3^{\text{rd}}$ layer}
    \label{fig:layer3_Tinyimagenet}
  \end{subfigure}
  \hspace{0.5cm}
  \begin{subfigure}{0.45\linewidth}
    \includegraphics[width=\linewidth]{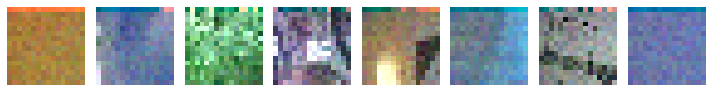}  
    \caption{$4^{\text{th}}$ layer}
    \label{fig:layer4_Tinyimagenet}
  \end{subfigure}
  \hspace{0.5cm}
  \begin{subfigure}{0.11\linewidth}
    \includegraphics[width=\linewidth]{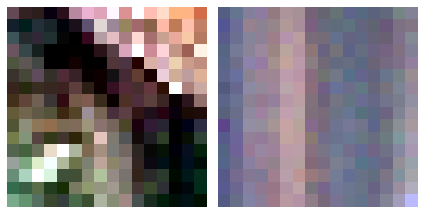}  
    \caption{$5^{\text{th}}$ layer}
    \label{fig:layer5_Tinyimagenet}
  \end{subfigure}
 \vspace{-0.2cm} \caption{Recovered patches from the first position using multiple adapter layers (TinyImageNet). None of the patches are recovered from the $1^{\text{st}}$ layer gradients. }
  \label{fig:recovery_layer_TinyImageNet}
\vspace{-0.1cm}\end{figure*}

\begin{figure*}
  \centering
 \begin{subfigure}{0.23\linewidth}
    \includegraphics[width=\linewidth]{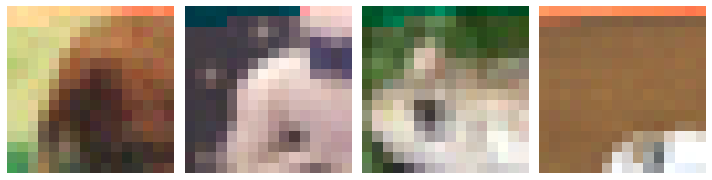}  
    \caption{$r=8$}
    \label{fig:r=8_CIFAR10}
  \end{subfigure}\\
   \hspace{0.5cm}
  \begin{subfigure}{0.95\linewidth}
    \includegraphics[width=\linewidth]{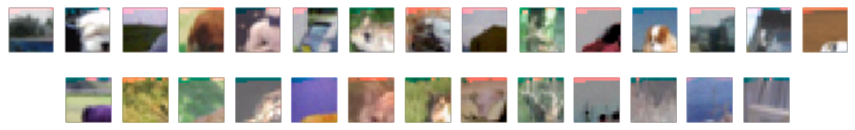}  
    \caption{$r=64$}
    \label{fig:r=64_CIFAR10}
  \end{subfigure}
  %\hspace{0.5cm}  
 \vspace{-0.2cm} \caption{Impact of bottleneck dimension $r$ on patch reconstruction (CIFAR-10). }
  \label{fig:r_CIFAR10}
\vspace{-0.1cm}\end{figure*}

\begin{figure*}
  \centering
 \begin{subfigure}{0.23\linewidth}
    \includegraphics[width=\linewidth]{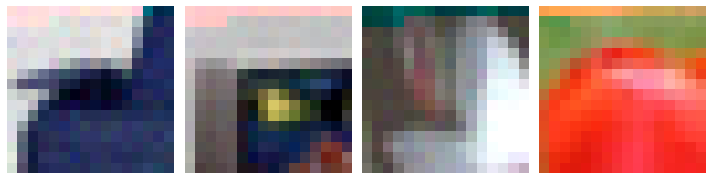} 
    \caption{$r=8$}
    \label{fig:r=8_CIFAR100}
  \end{subfigure}\\
   \hspace{0.5cm}
  \begin{subfigure}{0.97\linewidth}
    \includegraphics[width=\linewidth]{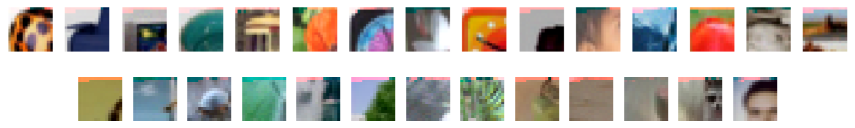}  
    \caption{$r=64$}
    \label{fig:r=64_CIFAR100}
  \end{subfigure}
  %\hspace{0.5cm}  
 \vspace{-0.2cm} \caption{Impact of bottleneck dimension $r$ on patch reconstruction (CIFAR-100). }
  \label{fig:r_CIFAR100}
\vspace{-0.1cm}\end{figure*}

\begin{figure*}
  \centering
 \begin{subfigure}{0.13\linewidth}
    \includegraphics[width=\linewidth]{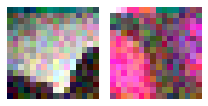}  
    \caption{$r=8$}
    \label{fig:r=8_TinyImageNet}
  \end{subfigure}\\
   \hspace{0.5cm}
  \begin{subfigure}{0.94\linewidth}
    \includegraphics[width=\linewidth]{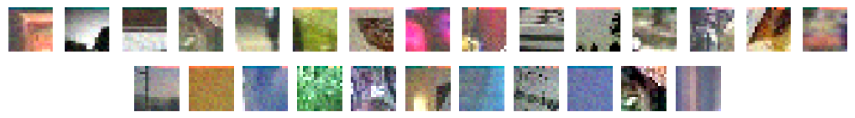}  
    \caption{$r=64$}
    \label{fig:r=64_TinyImageNet}
  \end{subfigure}
  %\hspace{0.5cm}  
 \vspace{-0.2cm} \caption{Impact of bottleneck dimension $r$ on patch reconstruction (TinyImageNet). }
  \label{fig:r_TinyImageNet}
\vspace{-0.1cm}\end{figure*}

\begin{figure*}
  \centering
  % Subfigure 1
    \begin{subfigure}{0.3\linewidth}
    \includegraphics[width=\linewidth]{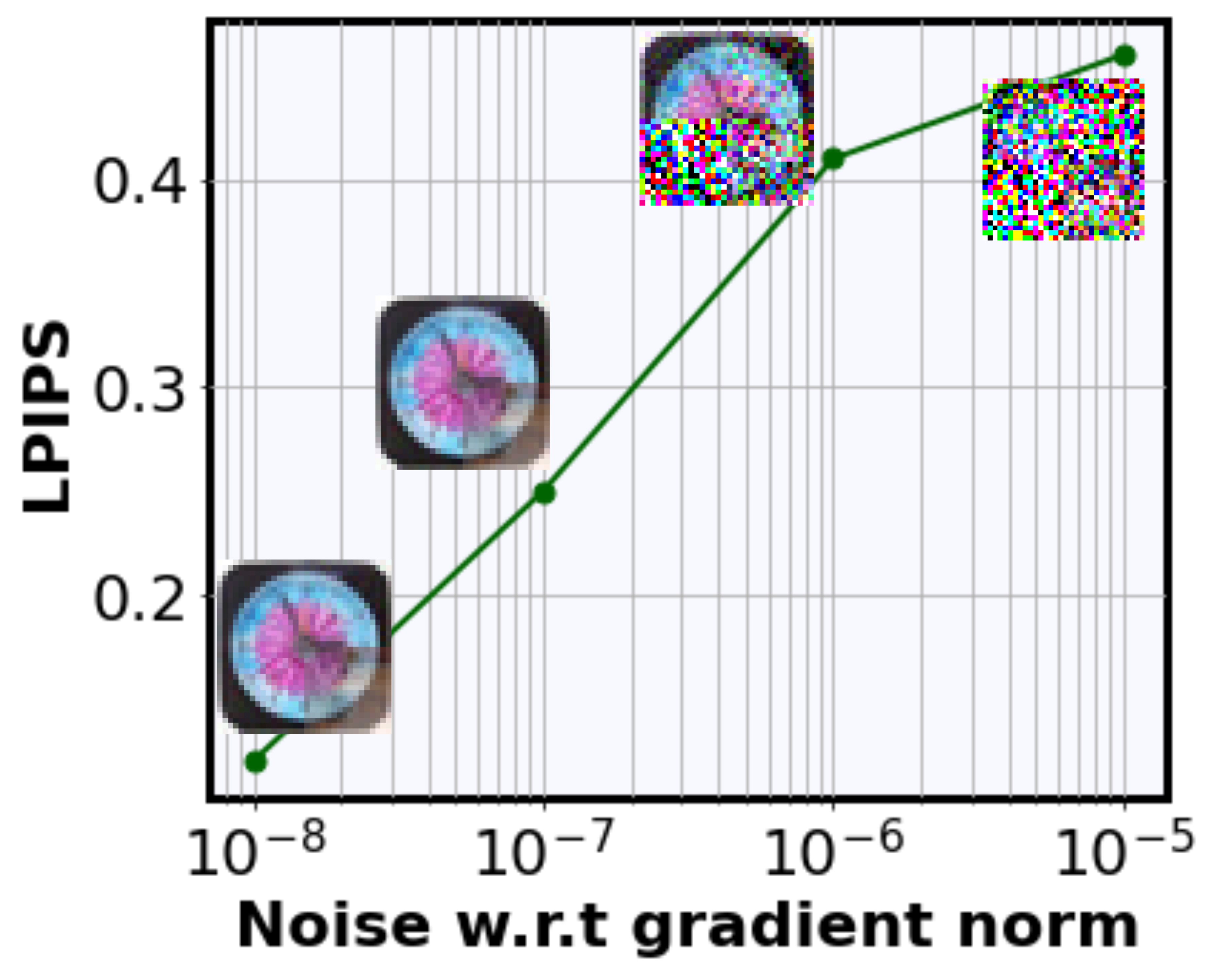}  
    \caption{Noise}
    \label{fig:noise}
  \end{subfigure} 
   \begin{subfigure}{0.3\linewidth}
    \includegraphics[width=\linewidth]{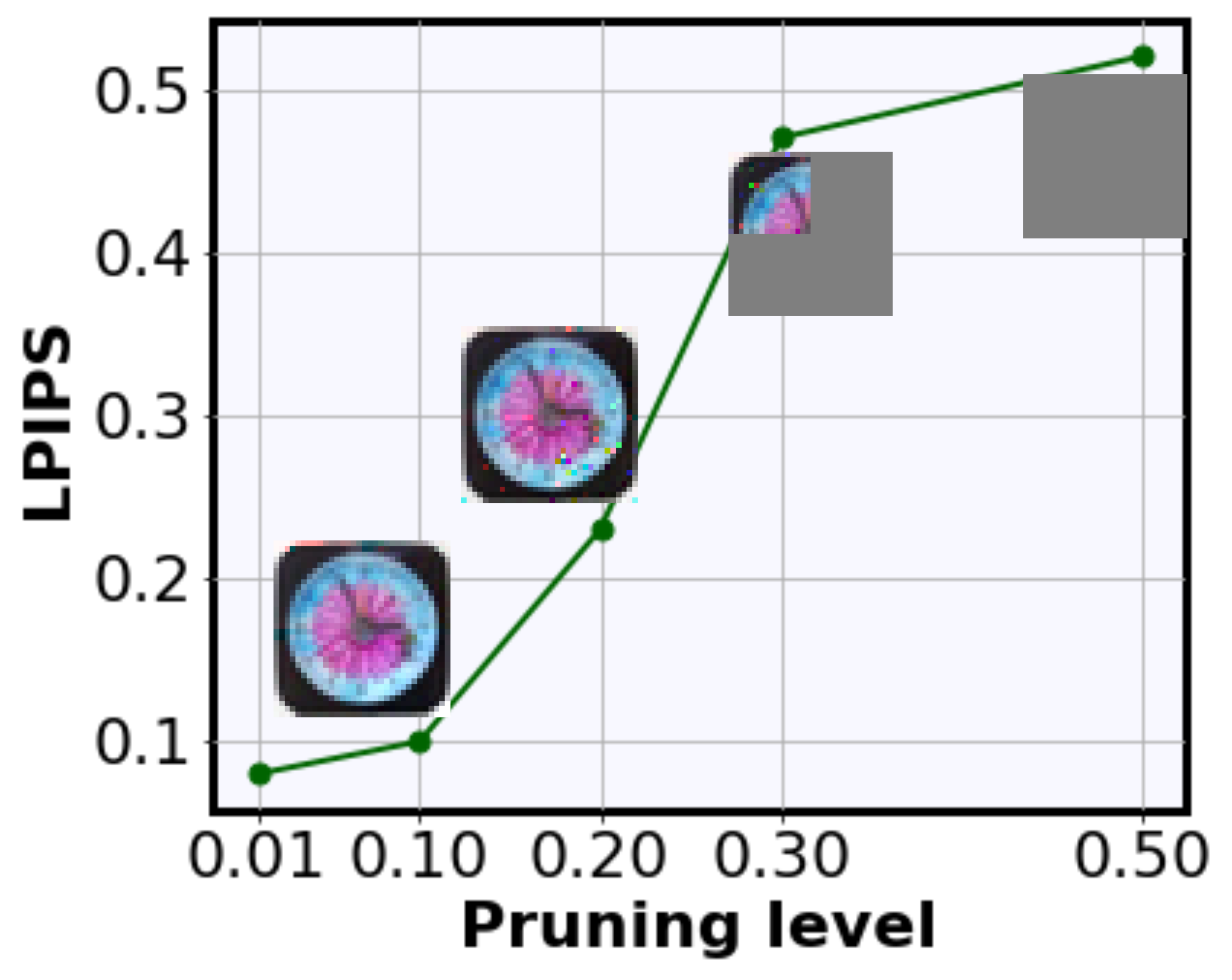}  % Insert your first image here
    \caption{Pruning}
    \label{fig:pruning}
  \end{subfigure} 
  %\hspace{0.5cm} 
  \begin{subfigure}{0.295\linewidth}
    \includegraphics[width=\linewidth]{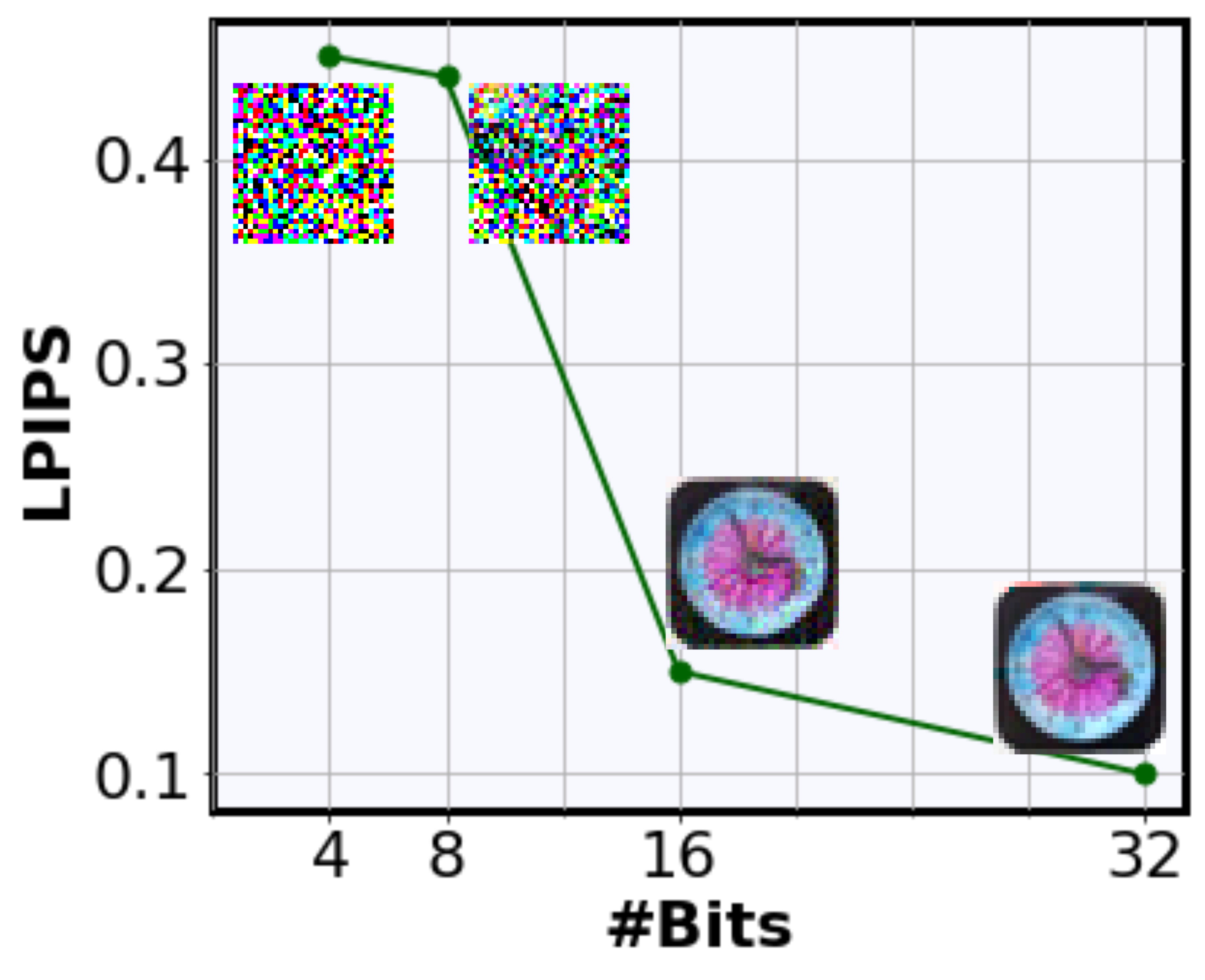}  
    \caption{Quantization}
    \label{fig:quant}
  \end{subfigure}
   \vspace{-0.2cm} \caption{Performance against mitigation strategies (CIFAR-100, batch size $32$).  Lower LPIPS denotes better reconstruction. 
   }
  \label{fig:plots_defense}
\end{figure*}

\begin{figure*}
  \centering 
   \begin{subfigure}{0.43\linewidth}
    \includegraphics[width=\linewidth]{figures/CIFAR100_Ground_truth.png} 
    \caption{Original images}
    \label{fig:original_CIFAR100_FedAvg}
  \end{subfigure}
  \hspace{0.5cm}
  \begin{subfigure}{0.43\linewidth}
    \includegraphics[width=\linewidth]{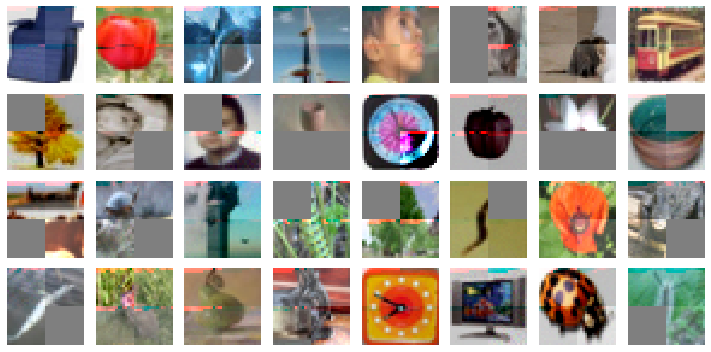}  
    \caption{Recovered}
    \label{fig:rec_CIFAR100_FedAvg}
  \end{subfigure}
\vspace{-0.2cm}  \caption{Recovered images for FedAvg with $5$ local training rounds (CIFAR-100). }
  \label{fig:recovery_CIFAR100_FedAvg}
\vspace{-0.3cm}\end{figure*}

\begin{figure*}
  \centering
  % Subfigure 1
  %  \begin{subfigure}{0.45\linewidth}
  %   \includegraphics[width=\linewidth]{figures/CIFAR100_Ground_truth.png}  
  %   \caption{Original images}
  %   \label{fig:original_opt_CIFAR100}
  % \end{subfigure}
  % \hspace{0.5cm}
  \begin{subfigure}{0.43\linewidth}
    \includegraphics[width=\linewidth]{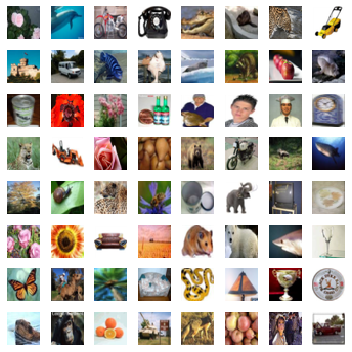}  
    \caption{Original images}
    \label{fig:GT_64images_CIFAR100}
  \end{subfigure}
  \hspace{0.5cm}
  \begin{subfigure}{0.43\linewidth}
    \includegraphics[width=\linewidth]{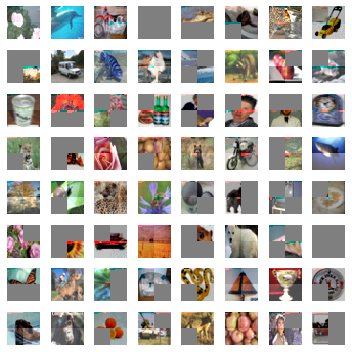}  
    \caption{Recovered (\name)}
    \label{fig:rec_64images_CIFAR100}
  \end{subfigure}
 \vspace{-0.2cm} \caption{Recovered images for a batch of size $64$ (CIFAR-100). }
  \label{fig:recovery_64images_CIFAR100}
\vspace{-0.1cm}\end{figure*}

\begin{figure*}[h]
  \centering
  % Subfigure 1
   \begin{subfigure}{0.4\linewidth}
    \includegraphics[width=\linewidth]{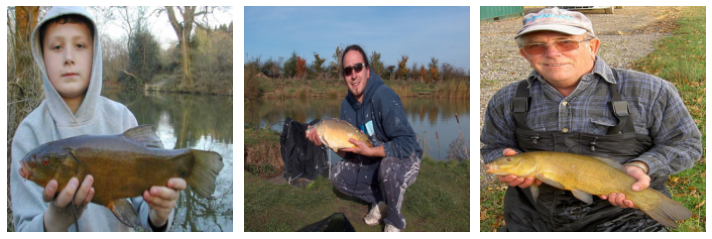}  % Insert your first image here
    \caption{Ground-truth}
    \label{fig:GT}
  \end{subfigure}
  \hspace{0.3cm}
  \begin{subfigure}{0.4\linewidth}
    \includegraphics[width=\linewidth]{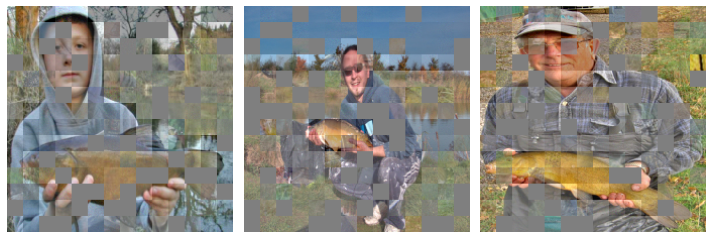}  
    \caption{Recovered}
    \label{fig:rec}
  \end{subfigure}
   \vspace{-0.3cm} \caption{Recovered images (ImageNet). }
  \label{fig:imagenet}
\vspace{-0.3cm}\end{figure*}

%\noindent
%{\bf Reconstruction of patches from multiple adapter layers. }

\subsection{Different Model Architectures}
Table \ref{table:arch} shows our results for ViT-L/16 and ViT-B/32 with a batch size of $32$.  
% We observed an interesting trade-off. 
We observed that for a fixed embedding dimension $D$, more encoders (ViT-L/16) can speed up our attack.  
ViT-L/16 ($24$ encoders) recovers an image in just $2$ rounds, compared to $4$ rounds for ViT-B/16 ($12$ encoders). 
When the number of encoders is fixed, we observed an interesting relation between  $D$ and patch size $P$.
% attack improves with a larger $D$.   
In ViT-B/32, each $(P,P)=(32,32)$ patch flattens to a $P^2C=3072$-dimensional vector ($C$ channels). 
If $D\geq P^2C$, as in ViT-B/16 ($P=16,D=768$) and ViT-L/16 ($P=16, D=1024)$, all pixels can be recovered. 
In ViT-B/32, $D<P^2C$, limiting  naive  recovery to  $D=768$ pixels. 
A simple solution is then to recover an average pixel from each $(2,2)$ region, yielding a lower resolution reconstruction. 
Fig. \ref{fig:arch} illustrates this for a recovered sample.

\subsection{Ablation Study}
{\bf Varying batch size. } We next demonstrate the reconstruction performance with varying batch size, bottleneck dimension and number of adapter layers for CIFAR-10 and TinyImageNet dataset (CIFAR-100 results were provided in Section \ref{sec:exp} in our main paper). In Figs. \ref{fig:CIFAR10_Batch} and \ref{fig:TinyImageNet_batch}, we observe that even for batch sizes as large as $64,96,128$, a notable amount of the patches are recovered. 

\noindent
{\bf Varying bottleneck dimension. } We next report the reconstruction rate for varying $r$, the bottleneck dimension within each adapter layer. Higher value of $r$ implies that more neurons are available in each adapter layer that can be leveraged for reconstruction. In Figs. \ref{fig:CIFAR10_r}, \ref{fig:TinyImageNet_r}, we observe that  as $r$ increases, more patches are recovered. 

%Finally,   
\noindent
{\bf Benefits of using multiple adapter layers. } For the experiments in Figs. \ref{fig:rec_CIFAR100}, \ref{fig:rec_TinyImageNet} and \ref{fig:rec_CIFAR10}, we have allocated $5$ adapter layers for the reconstruction of patches from each position. As mentioned in Section \ref{sec:exp}, images from CIFAR-10 and CIFAR-100 datasets are divided into $4$ patches. Therefore, for $4$ patches, we utilize $20$ adapter layers in total within a single training round. 
%In Fig. \ref{fig:CIFAR10_layer}, we observe that as we continue to utilize more adapter layers, more patches are recovered. 
For TinyImageNet, each image is divided into $16$ patches. The server aims to recover $4$ patches from $20$ adapter layers per training round. For this, the server sends malicious adapter parameters to recover patches from $4$ target positions by leveraging the adapter gradients received from the user in each round. Hence, all the patches are recovered over $4$ training rounds. In this regard, we next demonstrate the benefit of using multiple adapter layers in terms of attack success. In Figs. \ref{fig:CIFAR10_layer} and \ref{fig:TinyImageNet_layer}, we report the percentage of patches recovered per training round. As we observe, more patches are recovered as more adapter layers are being utilized. 

We further provide the illustration of the recovered patches in Figs. \ref{fig:recovery_layer_CIFAR10}-\ref{fig:r_TinyImageNet}. Figs. \ref{fig:recovery_layer_CIFAR10}, \ref{fig:recovery_layer_CIFAR100}, and \ref{fig:recovery_layer_TinyImageNet}, 
%we demonstrate the recovery of different patches from multiple layers. For this, 
demonstrate the recovery of the patches from the first position, i.e., top-left patch of the images from Figs. \ref{fig:original_CIFAR10}, \ref{fig:original_CIFAR100} and \ref{fig:original_TinyImageNet}. As described in Section \ref{sec:framework}, the weight and bias parameters in the adapter layers are designed such that patches from the target position can be recovered by leveraging the adapter gradients. Patches from all other positions will be filtered out by the activation function. We observe that by utilizing multiple adapter layers, we recover most of the target patches for this position. Moreover, in Figs. \ref{fig:r_CIFAR10}, \ref{fig:r_CIFAR100}, and \ref{fig:r_TinyImageNet}, we demonstrate the recovered patches from the same target position for $r=8$ in comparison with $r=64$. As we observe, more patches are retrieved from the adapter gradients when $r$ is increased from $8$ to $64$.

\subsection{Robustness Against Defense Mechanisms}
Fig.~\ref{fig:plots_defense} presents the attack performance against potential defense mechanisms, including noise addition \cite{Abadi2016}, pruning (top-$K$) \cite{Yujun2018, Alistarh2018} and stochastic quantization \cite{Dan2017}. Attack performance is measured in terms of average LPIPS score \cite{Zhang2018} between recovered and ground-truth images.  In Fig. \ref{fig:noise}, we vary the standard deviation of added Gaussian noise  with respect to the gradient norm.

\subsection{Attack to FedAvg}
We next consider the FedAvg setup, where each user performs multiple rounds of local training before sending the gradient to the server.  
We again leverage the activation structure from \cite{Fowl2022} (proposed for the FedAvg setting) in the down-projection layer within each adapter. 
At each global training round, each user performs local training for $5$ epochs, and sends the local gradient to the server. We demonstrate the reconstructed images in Fig. \ref{fig:rec_CIFAR100_FedAvg}, and observe that image patches can be recovered with high fidelity.

\subsection{Reconstruction on Additional Images}
%{\bf Recovery for a large batch size. } 
In Fig. \ref{fig:recovery_64images_CIFAR100}, we further demonstrate the reconstructed images from a larger batch size. 
For this, we consider the images from CIFAR-100 dataset for a batch of size $64$. As we observe in Fig. \ref{fig:recovery_64images_CIFAR100}, successful reconstruction of $75\%$  of the patches 
%for CIFAR-100 
%and $72\%$ for CIFAR-10) 
is obtained from the adapter gradients. 

\begin{table}[h]
\centering
 % \small
% \footnotesize
\scriptsize

 % \addtolength{\tabcolsep}{-0pt} 
\setlength{\tabcolsep}{3pt}
\begin{tabular}{@{}lc c c c c  @{}}\toprule

$\sigma$ &$1$ &$2$ & $3$  &$5$ &$10$\\

  \cmidrule{2-6}
%            & MSE   \\\midrule
 {\bf Gaussian} &$12$ & $30.4$ &$52.3$ &$77.3$ &$85.9$   \\
{\bf Laplacian} &$12.5$ & $35.1$ &$57.8$ &$70$ &$92.9$ \\
   \bottomrule
\end{tabular} 
% \addtolength{\tabcolsep}{2pt}

  % \vspace{-0.1cm}
 \vspace{-0.2cm}  \caption{$\%$ patches recovered with different $\sigma$ and distributions. } 
\label{table:pos}
 \vspace{-0.4cm}
\end{table}

\subsection{Attack Detectability} 
Our attack leverages the fact that users implicitly trust the server for the pretrained model and fine-tuning parameters. However, our malicious design may cause the users to question the integrity of the server. As described in Section \ref{sec:adapt}, to recover patches from a target position, our attack sets the weight rows to be identical in the first linear layer of the adapter modules. To make this design more stealthy, the server can introduce non-malicious weight rows and biases in-between. Moreover, for position encoding, any distribution can be used if they meet the criteria outlined in \eqref{eq:larger} and Section \ref{sec:LN1}.
Table \ref{table:pos} shows our results with lower standard deviation $\sigma$ across multiple distributions to improve stealth. Even with a $\sigma$ as small as $3$, our attack can recover $57.8\%$ of the patches (batch size $32$, CIFAR-100).

\subsection{Reconstructed Images from the ImageNet Dataset}
In Fig. \ref{fig:imagenet}, we present sample images from ImageNet. 

% As we observe, the images are retrieved with high fidelity. 


\end{document}